\pdfoutput=1
\documentclass[11pt,twoside,a4paper,cmspaper,final,collab]{cms-tdr}
\def\svnVersion{ccbfbb8}\def\svnDate{2025/10/31}\def\cmsCernNoTag{CERN-EP-2024-234}\def\cmsCernDate{\today}\def\cmsMessage{Published in the Journal of High Energy Physics as \href{http://dx.doi.org/10.1007/JHEP10(2025)219}{\doi{10.1007/JHEP10(2025)219}.}}
\begin{document}\cmsNoteHeader{SMP-22-008}

\providecommand{\cmsTable}[1]{\resizebox{\textwidth}{!}{#1}}
\providecommand{\cmsLeft}{left\xspace}
\providecommand{\cmsRight}{right\xspace}
\newcommand{\WW}{\ensuremath{\PW\PW}\xspace}
\newcommand{\PGnell}{\HepParticle{\PGn_{\!\ell}}{}{}\Xspace}
\newcommand{\ptrel}[1]{\ensuremath{p^\text{rel}_{\mathrm{T},\,#1}}\xspace}
\newcommand{\PTrel}[1]{\ensuremath{p^\text{rel}_{\mathrm{T}}(#1)}\xspace}
\newcommand{\etauh}{\ensuremath{\Pe\tauh}\Xspace} 
\newcommand{\mutauh}{\ensuremath{\PGm\tauh}\Xspace} 
\newcommand{\Nexp}{\ensuremath{N_\text{exp}}\Xspace} 
\newcommand{\Moone}{\ensuremath{m_{\circ 1}}\Xspace}

\cmsNoteHeader{SMP-22-008}
\title{Study of same-sign  \texorpdfstring{\PW}{W} boson scattering and anomalous couplings in events with one tau lepton from \texorpdfstring{{\Pp}{\Pp}}{pp} collisions at \texorpdfstring{$\sqrt{s} = 13\TeV$}{sqrt(s) = 13 TeV}}

\date{\today}

\abstract{
A first study is presented of the cross section for the scattering of same-sign \PW boson pairs via the detection of a \PGt lepton. The data from proton-proton collisions at the center-of-mass energy of 13\TeV were collected by the CMS detector at the LHC, and correspond to an integrated luminosity of 138\fbinv. Events were selected that contain two jets with large pseudorapidity and large invariant mass, one \PGt lepton, one light lepton (\Pe or \PGm), and significant missing transverse momentum. The measured cross section for electroweak same-sign \WW scattering is $1.44^{+0.63}_{-0.56}$ times the standard model prediction. In addition, a search is presented for the indirect effects of processes beyond the standard model via the effective field theory framework, in terms of dimension-6 and dimension-8 operators.
}

\hypersetup{
pdfauthor={CMS Collaboration},
pdftitle={Study of same-sign W boson scattering in events with one tau lepton from pp collisions at sqrt(s) = 13 TeV},
pdfsubject={CMS},
pdfkeywords={CMS, SMP, VBS, hadronic tau, machine learning, EFT}}

\maketitle 

\section{Introduction}\label{sec:intro}

The discovery of the Higgs boson~\cite{Khachatryan:2016vau} provided evidence that fermions and bosons acquire their masses through the Brout--Englert--Higgs mechanism of electroweak (EW) spontaneous symmetry breaking~\cite{Higgs:1964ia,Higgs:1964pj,PhysRevLett.13.321}. 
In this framework, vector boson scattering (VBS) processes play a special role. This occurs because in the standard model (SM) the unitarity of the scattering amplitude depends on a cancellation of diagrams involving the mediation of the Higgs boson between longitudinally polarized vector bosons. As a result, the contribution of the longitudinal polarization component to vector boson scattering is very small. Therefore, for even small deviations from the SM couplings of the Higgs boson to the vector bosons, the VBS cross section would diverge from the SM expectations with increasing center-of-mass energy. The measurement of VBS processes thus provides an indirect probe of physics beyond the SM (BSM), even for scenarios in which new resonances are not energetically accessible at the LHC~\cite{Chang:2013aya}.
The theoretical calculation of the effects of different sources of deviations from the SM is sensitive to the method chosen to ``unitarize'' the process at higher energies~\cite{Kilian:2014zja, PhysRevD.100.096003}.

\begin{figure}[htpb!]
    \centering
    \includegraphics[width=0.33\linewidth]{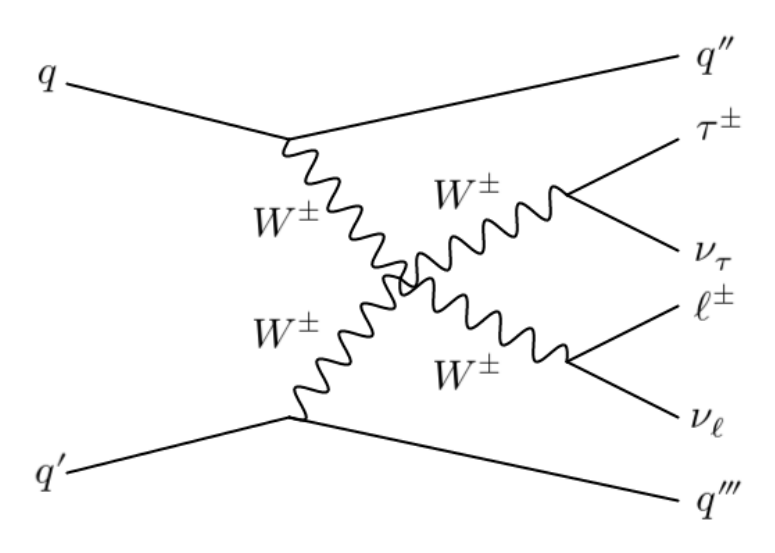}
    \includegraphics[width=0.33\linewidth]{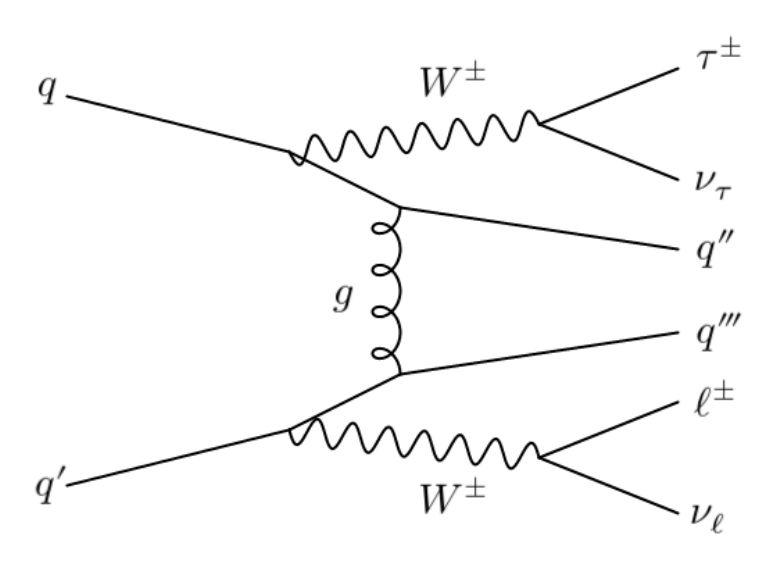}
    \caption{Representative tree-level Feynman diagrams contributing to the process $\PQq\PQq^\prime \to \PGt^{\pm}\PGnGt \Pell^{\pm} \nu_{\Pell} jj$, $\Pell=\Pe, \, \PGm$, leading to cross sections of order $\alpha_\text{EW}^6$ (\cmsLeft) and $\alpS^{2}\alpha_\text{EW}^4$ (\cmsRight).}
    \label{fig:feyndiag}
\end{figure}

At tree level, the topology commonly identified as VBS appears through purely EW diagrams of order $\alpha_\text{EW}^6$, where $\alpha_\text{EW} = g_W^2/4\pi$ is the electroweak SU(2) coupling. However, additional non-VBS electroweak diagrams in the same order also contribute, and the ensemble of all indistinguishable contributions constitutes the electroweak $\PW\PW$ production process. For simplicity, we refer to this entire production mechanism at order $\alpha_\text{EW}^6$ as ``EW VBS'' in the rest of this article.  Because it arises from higher-order electroweak interactions, EW VBS generally leads to a small cross section.
Along with these purely EW terms, experimental VBS signatures also include irreducible contributions that enter at order $\alpS^{2}\alpha_\text{EW}^4$, where \alpS is the strong coupling. These contributions are referred to as quantum chromodynamics (QCD) irreducible contributions. Typical Feynman diagrams for these processes are shown in Fig.~\ref{fig:feyndiag}.

The ATLAS collaboration provided the first evidence of same-sign \PW pair (SSWW) production via VBS in 2014 by studying final states with electrons and muons~\cite{PhysRevLett.113.141803}. The first observation, in the same final states, was presented by CMS in 2017~\cite{PhysRevLett.120.081801}.
Among the EW-mediated processes, SSWW EW VBS has the largest cross section ratio between the EW and QCD production modes~\cite{Covarelli:2021gyz}.

The study presented here of $\Pp\Pp\to\PWpm\PWpm+\text{jets}$ is based on data from proton-proton collisions at $\sqrt{s}=13\TeV$ collected by the CMS experiment at the CERN LHC from 2016 to 2018 and corresponding to an integrated luminosity of 138\fbinv.
We investigate a heretofore unexplored final state characterized by the decay of one of the scattered \PW bosons into a \PGt lepton that subsequently decays into hadrons (hadronic \PGt candidate, \tauh). The final state thus consists of a charged light lepton $\Pell = \Pe,\PGm$, the corresponding neutrino \PGnell, one \tauh candidate, the corresponding \PGnGt, and two jets produced by the quarks recoiling from the production of the \PW boson pair.
Contributions to this final state could arise from BSM scenarios that favor the \PGt over light leptons.
Examples of these scenarios are provided by the $SU(2)_{l}\times SU(2)_{h}\times U(1)$ family models, denoted in the following as $\textrm{G}(221)$ models, where the first two families of quarks and leptons transform according to the $SU(2)_{l}$ group and the third family according to the $SU(2)_{h}$ group~\cite{Chiang:2009kb, Hayreter:2019dzc, Calabrese:2021lcz}.
In these frameworks, a moderate mixing between the $\PW$ and \PWpr gauge fields suppresses the \PWpr-quark coupling while retaining a sizable $\PH \PW \PWpr$ coupling. In the context of the latest direct searches for $\PGt$ resonances performed by CMS~\cite{CMS:2018fza} and ATLAS~\cite{ATLAS:2024tzc}, this setup is not excluded for $m_{\PWpr}=3.0\TeV$ and can still yield an enhancement of the  $\Pp\Pp\to\Pell\PGnell\PGt\PGnGt+\text{jets}$ cross section.

The sensitivity to indirect BSM effects can be probed within the standard model effective field theory framework~\cite{BuchmuellerWyler, warsaw}. This theory is referred to in the literature as ``SMEFT'', and we adopt the formulation of Ref.~\cite{warsaw}. Assuming that new physics with energy scale $\Lambda_{\text{BSM}} \gg \Lambda_{\text{SM}}$ (where $\Lambda_{\text{SM}}$ is a characteristic energy value, such as the Higgs field vacuum expectation value) induces only perturbative effects in EW VBS processes, the theory is implemented by introducing the following effective Lagrangian:
\begin{equation}
    \mathcal{L}_{\text{eff}} = \mathcal{L}_{\text{SM}} + \sum_{D_\alpha>4}\sum_{\alpha}\frac{c_\alpha^{(D_\alpha)}}{\Lambda_{\text{BSM}}^{D_\alpha-4}}\mathcal{O}^{(D)}_\alpha,
\end{equation}
where the operators $\mathcal{O}_\alpha^{(D)}$ are constructed with SM fields at some dimension $D_\alpha$, and $c_\alpha^{(D_\alpha)}$ are the Wilson coefficients.
In this way, the contribution $\mathcal{A}_{\text{BSM}}$ to the total scattering amplitude due to the EFT operators is given by:
\begin{equation}
\begin{aligned}
    \abs{\mathcal{A}_{\text{BSM}}}^2 =& \sum_i^{D_i>4}\left[\frac{c_i^{(D_i)}}{\Lambda^{D_i-4}_{\text{BSM}}}2\operatorname{Re}\abs{\mathcal{A}_{\text{SM}}^* \mathcal{A}_{\mathcal{O}_i^{(D_i)}}} + \frac{{c_i^{(D_i)}}^2}{\Lambda^{2(D_i-4)}_{\text{BSM}}}\abs{\mathcal{A}_{\mathcal{O}_i^{(D_i)}}}^2\right] \\
    &+ \sum_{j\neq k}^{D_j,D_k>4}\frac{c_j^{(D_j)}c_k^{(D_k)}}{\Lambda^{D_j+D_k-8}_{\text{BSM}}}\operatorname{Re}\abs{\mathcal{A}^*_{\mathcal{O}_j^{(D_j)}}\mathcal{A}_{\mathcal{O}_k^{(D_k)}}}\label{eq:absm},
\end{aligned}
\end{equation}
where the first summation runs over the interference terms between SM and one $\mathcal{O}_i^{(D)}$ operator, the second one over the quadratic contributions of the operators, and the last one over interference between two different operators. The $\mathcal{O}_\alpha^{(D)}$, as well as the corresponding $c_\alpha^{(D_\alpha)}$, are classified according to their dimension $D_\alpha$ to provide a first categorization of their physics effects, as explained below.

A deviation from SM expectations in the observed data yields would result in non-zero Wilson coefficient values for operators sensitive to such deviations. This effect would thus provide guidance for characterizing BSM effects. Effective field theory interpretations of search outcomes have previously been presented by the ATLAS and CMS Collaborations~\cite{CMS:2013hdf, CMS:2016rtz, ATLAS:2016snd, ATLAS:2016nmw, ATLAS:2017pbb, CMS:2019uys, CMS:2019qfk, ATLAS:2021jgw, CMS:2022dmc, ATLAS:2023sua, ATLAS:2023dkz}, along with phenomenological reinterpretations of the electroweak VBS differential cross section measurements~\cite{Ethier:2021ydt}.
Assuming that the new operators still respect lepton-number conservation, those with odd dimensions are ruled out. As a consequence, in this study, we investigate operators of dimension-6 (dim-6) and dimension-8 (dim-8), which induce anomalous triple and quartic gauge couplings, modify boson-fermion couplings, and introduce fermion contact interactions~\cite{Eboli:2016kko, Falkowski:2016cxu, bellan2021}. We refer to any contribution of dimension greater than four as an EFT contribution.
Specifically, the dim-6 operators are implemented in the standard SMEFT framework via the Warsaw basis~\cite{warsaw}, while the dim-8 operators are introduced using the Eboli basis~\cite{Almeida:2020ylr}. 

In this paper, we introduce a machine-learning approach to identifying the single-\tauh final state in the SSWW VBS process and measuring the signal strength, defined as the ratio of the observed signal yield to that predicted by the model, for both the EW contribution (with the QCD contribution fixed) and for the combined EW+QCD combination of these processes (with the ratio between the EW and QCD contributions fixed). The same approach is implemented to develop models capable of discriminating possible EFT contributions from SM processes.

In the remainder of this paper, Section~\ref{sec:detector} describes the CMS detector, Section~\ref{sec:samples} presents the methods for simulating events, and Section~\ref{sec:obj} describes the particle reconstruction.  The selection of signal and control samples is described in Section~\ref{sec:strategy}, along with the background estimation, and in Section~\ref{sec:sys} we discuss systematic uncertainties. The results of the measurement are given in Section~\ref{sec:measurement}, and of the EFT interpretation in Section~\ref{sec:eft}. We conclude with a summary in Section~\ref{sec:summary}.

\section{The CMS detector}\label{sec:detector}
The central feature of the CMS apparatus is a superconducting solenoid of $6\unit{m}$ internal diameter, providing a magnetic field of 3.8\unit{T}. Within this magnetic field volume are a silicon pixel and silicon strip tracker, a lead tungstate crystal electromagnetic calorimeter (ECAL), and a brass-and-scintillator hadron calorimeter (HCAL), each composed of a barrel and two endcap sections. Forward calorimeters extend the coverage in pseudorapidity $\eta$ provided by the barrel and endcap detectors. Muons are detected in gas-ionization chambers embedded in the steel flux-return yoke outside the solenoid. A more detailed description of the CMS detector, together with a definition of the coordinate system used and the relevant kinematic variables, can be found in Ref.~\cite{CMS:2008xjf}.

The silicon tracker used in 2016 measured charged particles within the range $\abs{\eta} < 2.5$. For nonisolated particles of transverse momentum \pt in the range $1 < \pt < 10\GeV$ and $\abs{\eta} < 1.4$, the track resolutions were typically 1.5\% in \pt and 25--90 (45--150)\mum in the transverse (longitudinal) impact parameter. For isolated particles with $\pt = 100 \GeV$ emitted at $\abs{\eta} < 1.4$, the resolutions are approximately $2.8\%$ in \pt, and in impact parameter 10\mum (transverse) and 30\mum (longitudinal)~\cite{CMS:2014pgm}. At the start of 2017, a new pixel detector was installed~\cite{Phase1Pixel}; the upgraded tracker measured particles up to $\abs{\eta} < 3.0$ with typical resolutions of 1.5\% in \pt and 20--75\mum in the transverse impact parameter for nonisolated particles of $1 < \pt < 10\GeV$~\cite{DP-2020-049}.

Events of interest are selected using a two-tiered trigger system~\cite{CMStrig}. The first level, composed of custom hardware processors, uses information from the calorimeters and muon detectors to select events at a rate of around 100 kHz. The second level, known as the high-level trigger, consists of a farm of processors running a version of the full event reconstruction software, optimized for fast processing that reduces the event rate to around 1 kHz before data storage.

\section{Simulated samples}\label{sec:samples}
Monte Carlo (MC) simulation is used in the analysis for the design of the event selection, evaluation of signal efficiencies, and estimation of some backgrounds.
The EW VBS signal samples are simulated at leading order (LO) with six EW and zero QCD vertices with the \MGvATNLO v2.6.5 generator~\cite{Alwall:2014hca}, requiring a final state with $\Pell\PGnell, \PGt\PGnGt$ pairs from the decays of the two \PW bosons (Fig.~\ref{fig:feyndiag}, left). 
The \MGvATNLO generator is also used to simulate the QCD-mediated SSWW process, which is generated at LO with up to three additional partons in the matrix element calculations that have two QCD vertices at the tree level (Fig.~\ref{fig:feyndiag}, right).
The interference between the SSWW EW and QCD diagrams, including the terms of order $\alpS\alpha_\text{EW}^{5}$, contributes less than 4\% to the inclusive cross section for the EW signal over the phase space region of interest of the analysis and is therefore neglected.

A complete set of next-to-LO (NLO) QCD and EW corrections for the SSWW VBS processes, described above in the leptonic decay channel for each \PW boson, have been computed as a function of the invariant mass of the VBS jet system~\cite{CMS-PAPERS-SMP-19-012,Biedermann:2016yds,Biedermann:2017bss}. They reduce the LO cross section of the EW SSWW process by 10--15\%, with the correction increasing in magnitude with increasing dilepton and dijet invariant masses. 
This type of correction is applied only to the EW VBS process, and is not applied to the EFT modifications of the EW VBS processes.

The effects of the EFT dim-6 and dim-8 operators are simulated with \MGvATNLO at LO for the EW VBS processes. We do not include them in the background processes, as they would represent irrelevant contributions. For the dim-6 class, we introduce five bosonic operators acting on the scattering of the \PW bosons ($\mathcal{Q}_{W}$, $\mathcal{Q}_{HW}$, $\mathcal{Q}_{HWB}$, $\mathcal{Q}_{H\square}$, $\mathcal{Q}_{HD}$, where the second subscript in $\mathcal{Q}_{H\square}$ refers to the d'Alembertian that appears in the operator), two fermionic operators acting on contact interactions between fermions ($\mathcal{Q}_{ll}^{(1)}$, $\mathcal{Q}_{qq}^{(1)}$), and four mixed operators acting on interactions between massive bosons and fermions ($\mathcal{Q}_{Hl}^{(1)},  \mathcal{Q}_{Hq}^{(1)}$, $\mathcal{Q}_{Hl}^{(3)}$), defined in the Warsaw basis~\cite{warsaw} via the \textsc{SMEFTSim}~\cite{Brivio:2020onw,smeft1} package.
Some of the background processes of our study could be sensitive to the EFT dim-6 operators under investigation. However, these EFT background effects are better constrained in phase spaces orthogonal to the one covered by this analysis, and do not represent significant non-SM contributions.
For the dim-8 category, we introduce nine operators modifying the interaction between two scattering \PW bosons ($\mathcal{O}_{S0}$, $\mathcal{O}_{S1}$, $\mathcal{O}_{M0}$, $\mathcal{O}_{M1}$, $\mathcal{O}_{M7}$, $\mathcal{O}_{T0}$, $\mathcal{O}_{T1}$, $\mathcal{O}_{T2}$ defined in the Eboli basis~\cite{Almeida:2020ylr, Eboli:2016kko}. In both cases, the effects of the new EFT operators are evaluated and stored using the \MGvATNLO reweighting technique~\cite{reweight2016}, and no LO-to-NLO correction is imposed on their contribution.

The \POWHEG v2 generator~\cite{Nason:2004rx, Frixione:2007vw, Alioli:2010xd} is used to simulate, at NLO accuracy in QCD, the following processes: production of \ttbar pairs in which each \PQt quark decays to a \PQb quark and a lepton pair (dileptonic \ttbar); $\PQt\PW$; Higgs boson production mediated by gluon-gluon and vector-boson fusion; and diboson production.

Production of $\ttbar\PW$, $\ttbar\PZ$, $\ttbar\PGg$, triple vector boson, vector boson associated with a Higgs boson, and Drell--Yan background events are simulated at NLO accuracy in QCD using the \MGvATNLO generator. The $\PQt\PZ\PQq$ process is simulated at NLO in the four-flavor scheme using \MGvATNLO.
We generally refer to $\PZ\PZ$, $\PZ\PGg$, $\PW\PGg$, $\PW\PZ$, tribosons, associated production of a quark-antiquark top pair with a \PGg, \PZ, or \PW boson, and $\PQt\PZ\PQq$ processes as ``others''. 
Higgs boson production via gluon-gluon fusion and vector-boson fusion, as well as the production of a vector boson in association with a Higgs boson, are collectively referred to as ``opposite sign'' (OS) since they, alongside DY, contribute to the final event selection when one of the leptons has a misreconstructed charge. OS and DY contributions are treated coherently in the statistical fits described in Sections~\ref{sec:measurement} and~\ref{sec:eft}.

The NNPDF3.1 next-to-NLO~\cite{thennpdfcollaboration2017parton} parton distribution functions (PDFs) are used in the simulation of the background and signal samples.
The generators used for signal and background processes are interfaced with the \PYTHIA 8.306~\cite{Bierlich:2022pfr} program, with the CP5 tune~\cite{Sirunyan_2020}, to model parton showering and hadronization.

Additional collisions in the same or adjacent bunch crossings (pileup) are included by superimposing simulated minimum bias interactions onto the hard-scattering process, with a multiplicity distribution matching the one that is observed in the data. Simulated events are propagated through the full \GEANTfour-based simulation~\cite{Agostinelli:2002hh} of the CMS detector.

\section{Event reconstruction}\label{sec:obj}
Events are selected for the signal measurement and the estimation of most of the backgrounds that have passed a trigger requiring the presence of either: one muon with $\pt > 24\GeV$ in 2016 and 2018 or $\pt > 27\GeV$ in 2017; one electron with $\pt > 27\GeV$ in 2016 or $\pt > 32\GeV$ in 2017 and 2018.
Data are selected for the additional control samples used for background estimation that were recorded with triggers requiring the presence of either a jet with $\pt > 40 \GeV$ or $\HT > 350\GeV$, where $\HT$ is the scalar sum of the jet transverse momenta in an event.

Particle candidates are processed with an optimized combination of all subdetector information using the CMS particle-flow (PF) algorithm~\cite{CMS:2017yfk} that reconstructs and identifies each individual particle in the event.
The missing transverse momentum vector \ptvecmiss is defined as the projection onto the plane perpendicular to the beam axis of the negative vector sum of the momenta of all reconstructed PF objects in an event. Its magnitude is referred to as \ptmiss.

Jets are reconstructed by clustering PF candidates using the anti-\kt jet finding algorithm~\cite{Cacciari:2008gp, Cacciari:2011ma} with a distance parameter of 0.4. Jets are calibrated in the simulation, and separately in data, accounting
for energy deposits of neutral particles from pileup and any nonlinear detector response.
The effect of pileup is mitigated through a charged-hadron subtraction technique~\cite{Cacciari:2007fd} that removes the energy of charged hadrons not originating from the primary vertex (PV) of the event. Corrections to jet energies to account for the detector response are propagated to \ptmiss.
Jets are required to have $\pt>30\GeV$, $\abs{\eta}<5$, and to meet jet quality criteria with measured efficiencies that are almost 100\%, for both data and simulated samples~\cite{CMS-PAS-JME-10-003,CMS-PAS-JME-16-003}.

The PV is taken to be the reconstructed vertex with the largest value of summed physics-object $\pt^2$, as described in Section 9.4.1 of Ref.~\cite{CMS-TDR-15-02}. 

A deep neural network-based tagger, \textsc{DeepJet}~\cite{Bols:2020bkb, CMS:2017wtu, CMS:2018lgn}, is used to identify jets stemming from the hadronization of \PQb quarks, utilizing information from the tracks, neutral particles, and the secondary vertices within the jet.  The efficiency and purity of the resulting ``\PQb jets'' are classified in terms of various working points. The analysis uses a medium working point that correctly identifies \PQb jets with an efficiency of about 70\%, and a loose working point with an efficiency of about 85\%. The misidentification rates for gluon or light-flavor quark jets for these two working points are 1.0\% and 10\%, respectivsely.

Electrons (muons) are reconstructed by associating a track reconstructed in the tracking detectors with a cluster of energy in the ECAL (track in the muon system). The candidates are required to originate from the PV, pass quality selection criteria, and  be isolated from other activity in the event.
For this purpose we define a lepton relative isolation variable $I_{\text{rel}}$ based on the energy deposited on a cone $\Delta R = \sqrt{\smash[b]{(\Delta\eta)^2+(\Delta\phi)^2}} = 0.3 \ (0.4)$ around the electron (muon).  Specifically, $I_{\text{rel}}=(E_{\text{ch}}+E_{\text{nh}}+E_{\text{ph}}-0.5\,E_{\text{ch}}^{\text{PU}})/\pt$, where $E_{\text{ch}}$ is the transverse energy deposited by charged hadrons from the PV, $E_{\text{nh}}$ and $E_{\text{ph}}$ are the transverse energies of the neutral hadrons and photons, respectively, and \pt is the electron or muon transverse momentum.  The term $0.5\,E_{\text{ch}}^{\text{PU}}$ accounts for the contribution of neutral particles from pileup vertices, taken as half the energy of the charged particles from pileup vertices. 

The quality criteria for the light lepton selection are based on the isolation from other particles in the event, the impact parameter of the candidate with respect to the PV, and the electromagnetic shower shape in the electron case. These are implemented by their inclusion in multivariate discriminators.
In this analysis we make use of both ``loose'' and ``tight'' working points of these discriminators.  The efficiency for loose (tight) electrons is 90\,(80)\%~\cite{CMS:2020uim}.  The corresponding efficiencies for muons are 99\,(95)\%~\cite{CMS:2018rym}.

For a loose electron (muon) we require $\pt > 15\GeV$ and $\abs{\eta} < 2.5\, (2.4)$ and $I_{\text{rel}}^{\text{electron}} < 0.20$ ($I_{\text{rel}}^{\text{muon}} < 0.40$).  For a tight electron (muon), the criteria are $\pt> 30\GeV$ (35\GeV for electrons in 2017 and 2018), and $I_{\text{rel}}^{\text{electron}} < 0.08$ ($I_{\text{rel}}^{\text{muon}} < 0.15$).

 Hadronically decaying \PGt leptons \tauh are reconstructed from jets using the hadrons-plus-strips algorithm~\cite{CMS:2018jrd}, which combines 1 or 3 tracks with energy deposits in the calorimeters to identify the \PGt decay modes. Neutral pions are reconstructed as strips with dynamic size in the $(\eta,\phi)$ plane from reconstructed electrons and photons, where the strip size varies as a function of the \pt of the electron or photon candidate.
To further distinguish genuine \tauh decays from jets originating from the hadronization of quarks or gluons, and from electrons or muons, we make use of the \textsc{DeepTau} algorithm~\cite{CMS:2022prd}. Information from all individual reconstructed particles near the \tauh axis is combined with properties of the \tauh candidate and the event to provide separate discriminators against hadronic jets ($D_j$), electrons ($D_e$), and muons ($D_\mu$). Similarly to the selection of electrons and muons, we employ a loose set of criteria to select \tauh with $\pt > 30 \GeV, \abs{\eta} < 2.3$, and satisfying $D_j, D_e, D_\mu$ working points for which the genuine \tauh identification efficiencies are 70, 98, and 99.5\%, respectively. We also make use of a working point with a tighter $D_j$ threshold, for which the identification efficiency for genuine \tauh is 50\%.

\section{Analysis strategy and background estimation}\label{sec:strategy}

The analysis targets the EW VBS production of same-sign $\PW\PW$, with one of the \PW bosons decaying to a \PGt lepton and the other into a \PGm or an \Pe, in association with two jets originating from the scattered incoming partons.
Events are first selected by requiring one electron or muon, one \tauh candidate, in each case satisfying the tight identification criteria, no additional loose leptons (\Pe, \PGm, or \tauh candidates), and at least two jets with a pseudorapidity separation $\abs{\Delta\eta}>2.5$. Among all the possible jet pairs that satisfy the latter requirement, the pair with the highest invariant mass $m_{jj}$ is chosen. We further define the signal region (SR) and several disjoint control regions (CRs) to estimate and validate the background predictions, as specified in the following paragraphs.

The SR is designed to enhance the yield of the EW VBS signal while minimizing that of the background. Events with a same-sign $\Pell\tauh$ pair, $\ptmiss>50\GeV$, and $m_{jj}>500\GeV$ are selected.
In this region, almost 95\% of the background events are contain nonprompt lepton candidates, which arise from jets misreconstructed as \Pe, \PGm, or \tauh, including genuine leptons from the decays of hadrons within jets.  About 2\% of background events arise from $\PZ/\PGg^*\!+\text{jets}$ and 1\% from dileptonic \ttbar production. Including both the $\etauh$ and $\mutauh$ channels, the SR comprises a total of 1047 observed events, with an expected 30 EW SSWW VBS events, 3 QCD-mediated SSWW VBS events, and 998 background events.

Nonprompt leptons are produced mainly by QCD-mediated multijet, associated $\PW+\text{jets}$, and hadronic and semileptonic \ttbar production.  They are estimated from data CRs by the ``pass-fail'' method described in detail in Ref.~\cite{fake2014}.
This method estimates the probability that a nonprompt lepton passes the tight selection criteria by using a region depleted of prompt leptons to determine transfer factors, functions of their \pt and $\eta$.  These are then used to calculate the nonprompt contributions in the main SRs and CRs, applying them to auxiliary regions defined as the SRs and main CRs, but requiring that at least one of the light lepton and \tauh pass the loose selection, while failing the tight one. 

\begin{figure}[htpb!]
    \centering
    \includegraphics[width=0.49\linewidth]{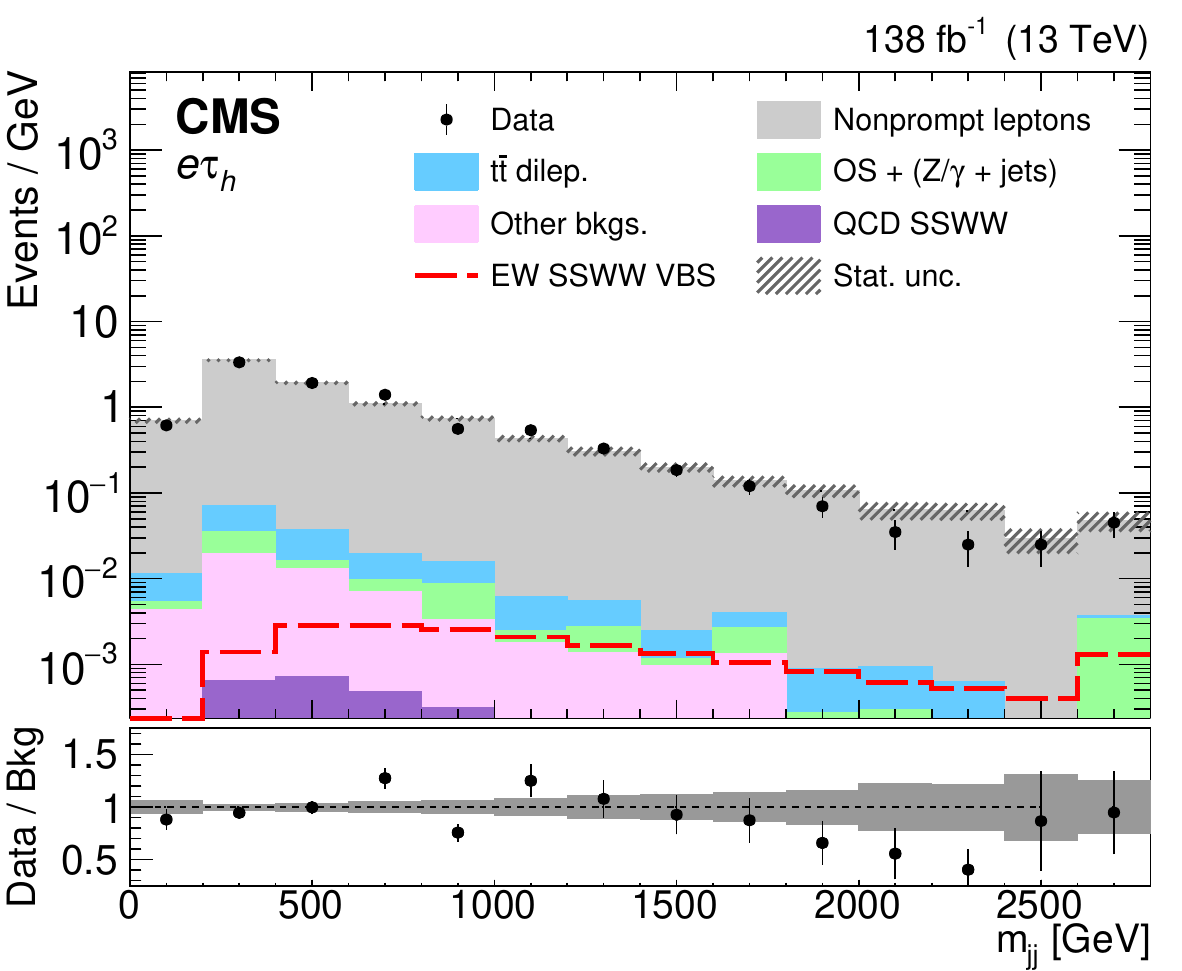}
    \includegraphics[width=0.49\linewidth]{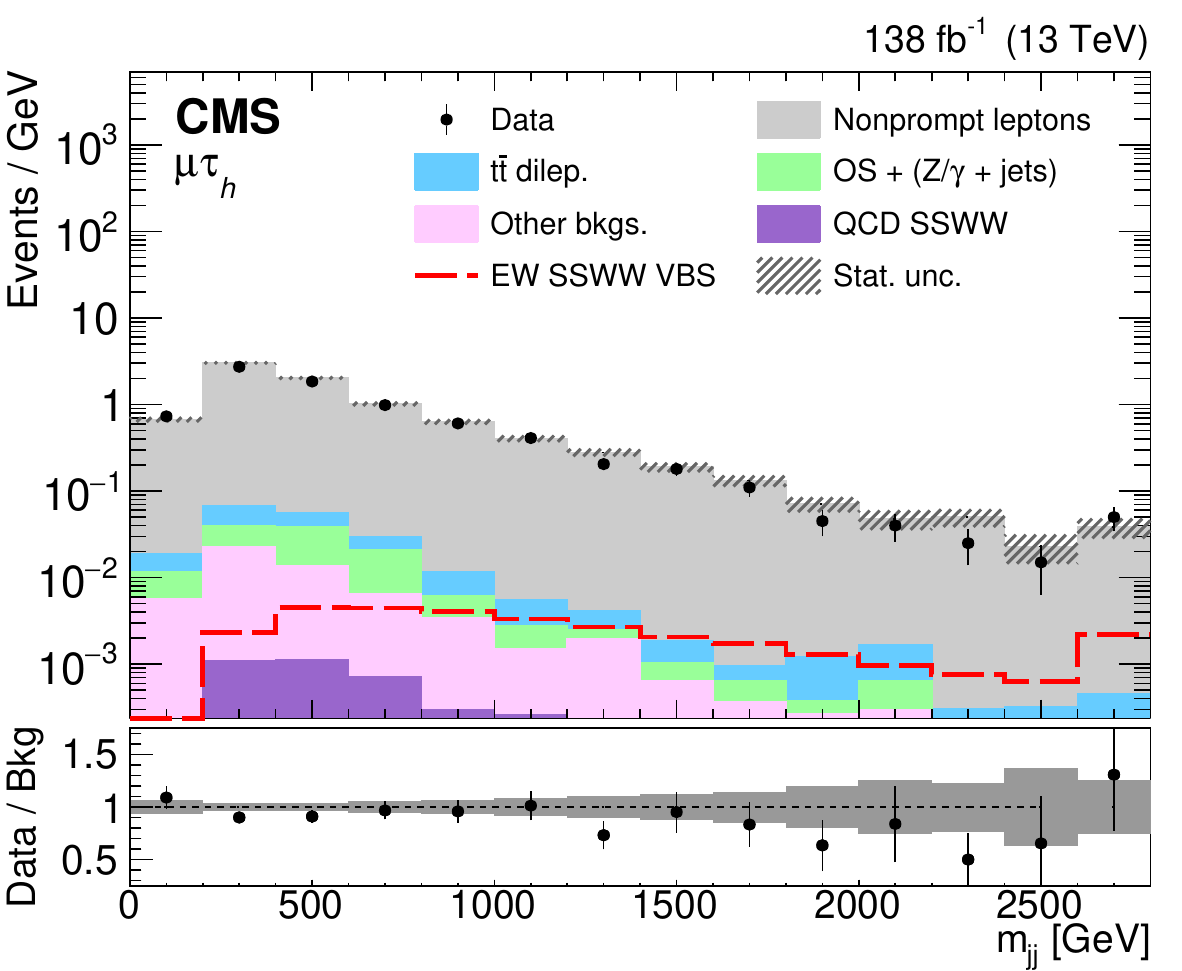}
    \caption{Distributions in the invariant mass of the dijet system for the data and the pre-fit background prediction for the (\cmsLeft) \etauh and (\cmsRight) \mutauh nonprompt VRs. The stacked filled histograms show the background components, with the overflow count included in the last bin. The expectation for the EW SSWW signal is shown by the dashed red line.  The hatched error band shows the bin-by-bin statistical uncertainty. The lower panels show the ratio of data to the total background prediction, with statistical uncertainties indicated by error bars and shading, respectively. In all the panels, the vertical bars represent the statistical uncertainty assigned to the observed number of events.}
    \label{fig:fake_val}
\end{figure}

For this background source, we define a ``QCD-enriched'' CR and a ``nonprompt'' validation region (VR). To define both the CR and the VR, the transverse mass $\mT(\Pell,\ptmiss)$ of the system comprising the light lepton and \ptmiss is introduced as follows:
\begin{equation}
  \mT(\Pell, \ptmiss) = \sqrt{\smash[b]{2\pt^{\Pell} \ptmiss [1-\cos{\Delta\phi}]}},\label{eq:mT}
\end{equation}
where $\Delta\phi$ is the azimuthal separation between the lepton momentum vector and \ptvecmiss.
The QCD-enriched CR is used to perform the first step of the nonprompt-lepton background estimations; it contains events with only one loosely identified lepton (\Pe, \PGm, \tauh), $\ptmiss< 50\GeV$, and $\mT(\Pell, \ptmiss)<50\GeV$ selected from data collected with a jet-based trigger.
The nonprompt VR serves to validate the yield estimate from the pass-fail method; it contains events with an SS $\Pell\tauh$ pair and $\ptmiss< 50$\GeV passing the same lepton-based triggers mentioned in Section~\ref{sec:obj}.  Lepton candidates in this VR arise mainly from $\PW+\text{jets}$ and QCD multijet production.
The orthogonality of the QCD-enriched CR and nonprompt VR is ensured by the fact that the triggers used to define them do not overlap.
The data and estimated background $m_{jj}$ distributions are compared in the nonprompt VRs for the electron and muon final states in Fig.~\ref{fig:fake_val}.
The background yields are evaluated before (``pre-fit'') the maximum likelihood (ML) fit introduced at the end of this section. The plots show that the data generally agree with the prediction within uncertainties.

Finally, we define \ttbar and OS/DY CRs to constrain the MC simulations of these background sources.
Events with an OS $\Pell\tauh$ pair and no loose \PQb-tagged jets are selected for the OS/DY CR; events with an OS $\Pell\tauh$ pair, at least one ``medium'' \PQb-tagged jet, and $\ptmiss> 50\GeV$ are selected for the \ttbar CR.
Both the \ttbar and OS/DY CR are included in the simultaneous ML fit along with the SR. The normalization of the dileptonic \ttbar is allowed to freely float. For the OS+DY background, we apply a normalization uncertainty correlated between the OS/DY CR and the SR, as described in Section~\ref{sec:sys}.
For the \ttbar normalization, the post-fit results agree with the pre-fit ones, as the observed value for the \ttbar normalization parameter is $1.01 \pm 0.04$.
A summary of the analysis phase space with the definitions of the SR and CRs is given in Table~\ref{tab:definition_SRCRs}, with the exclusion of the QCD-enriched CR.

\begin{table}[htpb!]
   \topcaption{Definitions of the SR, CRs, and VR defined for this analysis. The $\checkmark$ symbol indicates that the requirement described in the column heading is applied in that region, whereas the $\times$ symbol means that the opposite selection is applied. The symbols T and L refer to the tight and loose selection rules, respectively. The SR and three CRs (nonprompt, \ttbar, OS) are selected from an inclusive lepton trigger.}
  \label{tab:definition_SRCRs}
  \cmsTable{
  \begin{tabular}{lccccc}
    \hline
    \multirow{2}{*}{Region} & 1 T $\Pell$, 1 T \tauh, & $>=2$ jets & \multirow{2}{*}{SS $\Pell,\tauh$} & \multirow{2}{*}{$\ptmiss>$50\GeV} & \multirow{2}{*}{Additional requirements} \\
     & no L $\Pell$/\tauh & with $\abs{\Delta\eta}>2.5$ & & &\\
    \hline
    SR           & \checkmark & \checkmark & \checkmark & \checkmark & $m_{jj}>$500\GeV \\
    nonprompt VR & \checkmark & \checkmark & \checkmark & \scalebox{0.75}[1]{$\times$} & \\
    \ttbar CR    & \checkmark & \checkmark & \scalebox{0.75}[1]{$\times$} & \checkmark & \PQb-tagged jet (``medium'')\\
    OS/DY CR        & \checkmark & \checkmark & \scalebox{0.75}[1]{$\times$} & \checkmark & \PQb-tagged jet veto (``loose'') \\
    \hline
  \end{tabular}
  }
\end{table}

Because of the large background and complex signal topology, sets of significant features to separate signals and backgrounds are combined in three machine-learning discriminators, each targeting a different signal.  The discriminators are the outputs of feed-forward deep neural networks (DNNs). The three DNN models are devised to separate the SM EW VBS (SM DNN), EFT dim-6 (dim-6 DNN), and EFT dim-8 (dim-8 DNN) from the SM background processes.
In particular, in order to optimize the range of sensitivity of each of the EFT DNN models, we use a balanced mixture (i.e., an equal combination) of contributions from different EFT operators, which include both linear and quadratic terms, as the signal sample. This allows us to avoid the complexities associated with the use of several single-operator models, while preserving discrimination power across the various operators to which this analysis is mostly sensitive.
For the first EFT DNN model the signal is represented by a mixture of EFT linear and quadratic contributions weighted with unity values for $c_{HW}$ and $c_W$. For the second model a mixture of linear and quadratic contributions from all the dim-8 EFT operators included in the study is considered, with the corresponding Wilson coefficient values approximately equal to the 95\% confidence level (\CL) limits obtained performing a statistical fit with \Moone (defined in Eq.~\ref{eq:mo1} below) with the strategy outlined in Section~\ref{sec:eft}. In both cases, the SM EW VBS contribution is considered part of the background sample. The optimized models consist of 1, 2, and 4 hidden layers. Adam Optimizer~\cite{kingma2017adam}, early stopping, dropout, and L2 regularization~\cite{Goodfellow-et-al-2016} techniques are utilized to avoid the overfitting effect, consisting of excessive adaptation to training data of the DNN model.

\begin{figure}[htpb!]
    \centering
    \includegraphics[width=0.49\linewidth]{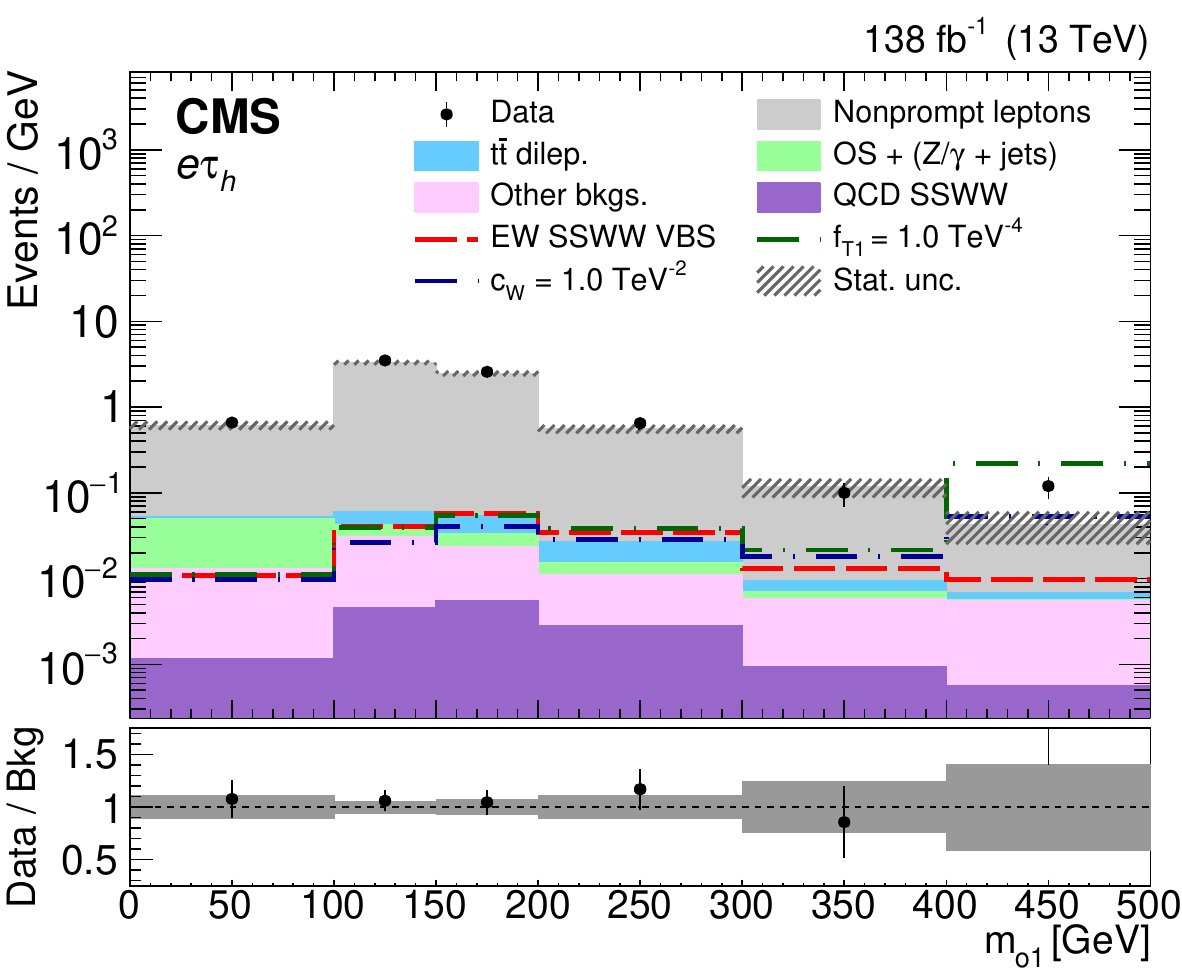}
    \includegraphics[width=0.49\linewidth]{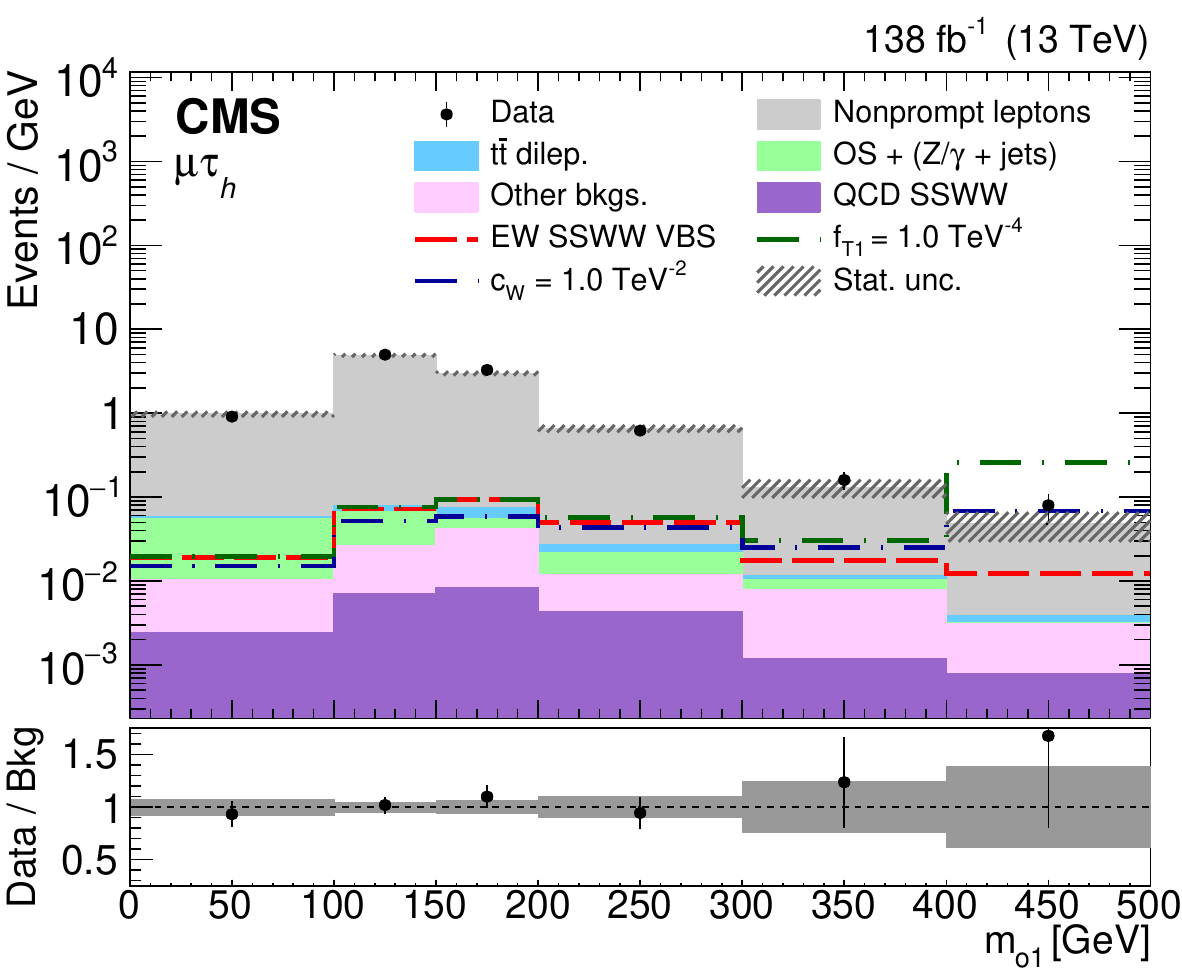}
    \caption{Distributions in \Moone transverse mass for the data and the pre-fit background prediction for the (\cmsLeft) \etauh and (\cmsRight) \mutauh SRs. The stacked filled histograms show the background components, and the overflow count is included in the last bin. The expectations for the EW SSWW signal, the $\mathcal{O}_{W}$ dim-6 operator with $c_{W}=1\TeV^{-2}$, and the $\mathcal{Q}_{T1}$ dim-8 operator with $f_{T1}=1\TeV^{-4}$ are shown by the dashed red, blue, and green lines, respectively. For the latter two, the interference with SM and pure EFT contributions are summed together with the SM contribution. The hatched error band shows the bin-by-bin statistical uncertainty. The lower panels show the ratio of data to the total background prediction, with statistical uncertainties indicated by error bars and shading, respectively. In all the panels, the vertical bars represent the statistical uncertainty assigned to the observed number of events.}    
    \label{fig:mo1_sr}
\end{figure}

The sets of input variables for the DNNs, detailed in Table~\ref{tab:DNNfeatures}, are each constructed with a dedicated optimization that includes some quantities introduced to exploit the particular kinematical properties of the EW VBS SSWW reaction~\cite{trmasses}. The agreement of the input variables used to train the DNN models has been thoroughly checked in the CRs introduced at the beginning of this section. 
First, there are the transverse masses
\begin{align}
      m_{1\text{T}}^2 &= \left(\sqrt{m_{\PGt\Pell}^2 + {\pt^{\PGt\Pell}}^{2}} + \ptmiss\right)^2 - \abs{ \ptvec^{\PGt\Pell} + \ptvecmiss}^2,\ \text{and} \label{eq:m1T}\\
    \Moone^2 &= \left(\pt^{\PGt} + \pt^{\Pell} + \ptmiss\right)^2 - \abs{\ptvec^{\PGt} +\ptvec^{\Pell} + \ptvecmiss}^2.\label{eq:mo1}
\end{align}
The variable $m_{1\text{T}}$ is the transverse mass of the $\PGt\Pell$ system with \ptmiss.
For the second quantity, \Moone, the \PGt, \Pell momenta, and \ptmiss are treated as if coming from a system with a null invariant mass when calculating the transverse mass of the three objects. These two variables are a proxy for the energy of the scattering \PW boson pair, as well as for the angular distribution of the decay objects coming from that pair. With these quantities it is possible to access direct information on the process of interest for this study, more complete than the invariant mass of the VBS jet pair and the \pt of the leptons. In this way, it is possible to enhance the discrimination of the SSWW VBS signal processes against the background, especially when investigating the sensitivity to EFT contributions, as shown for \Moone in Fig.~\ref{fig:mo1_sr}. The three transverse masses introduced in Eqs.~\ref{eq:mT},~\ref{eq:m1T}, and~\ref{eq:mo1} are among the most important features for all of the DNNs. 

Furthermore, the transverse masses $\mT(\tauh, \ptvecmiss)$ and $\mT(\Pell+\tauh, \ptvecmiss)$, defined similarly to $\mT(\Pell, \ptvecmiss)$ in Eq.~\eqref{eq:mT}, are introduced. 
Next, we define the event Zeppenfeld variable~\cite{Rainwater:1996ud}:
\begin{equation}
    z_{\text{event}} = \frac{1}{2}\frac{\eta_{\Pell} - \abs{\frac{\eta_{j_1}+\eta_{j_2}}{2}}}{\eta_{j_1}-\eta_{j_2}} + \frac{1}{2}\frac{\eta_{\tauh} - \abs{\frac{\eta_{j_1}+\eta_{j_2}}{2}}}{\eta_{j_1}-\eta_{j_2}},\label{eq:zepp}
\end{equation}
introduced to exploit the centrality of the leptons in the EW VBS processes with respect to the scattered VBS jets $j_1, j_2$. Finally we add the component $\ptrel{\Pl j}$ of the \Pell or \tauh momentum $\vec{p}_{\Pl}$ perpendicular to the momentum $\vec{p}_{j}$ of VBS jet $j$:
\begin{equation}
    \ptrel{\Pl j} = \frac{\abs{\vec{p}_{\Pl}\times\vec{p}_{j}}}{\abs{\vec{p}_{j}}}.
\end{equation}
This variable evaluates how close a lepton is to the flight direction of a jet. For EW VBS processes, this quantity is expected to be distributed towards values larger than for the other processes, since the light lepton and the \tauh produced by the scattered \PW bosons are far from the VBS jets. The pre-fit \Moone distributions in the SRs for the electron and muon final states are shown in Fig.~\ref{fig:mo1_sr}.

\begin{table}[htpb!]
\centering
\topcaption{List of the input variables for the three DNN models developed in this study.  The check mark indicates that the variable is included in the DNN model identified in the column header.}
\begin{tabular}{cccc}
\hline
Input variable                              & SM DNN                    & dim-6 DNN                 & dim-8 DNN                 \\
\hline
\tauh \pt                                   & \checkmark & \checkmark & \checkmark \\
\Pell \pt                                   & \checkmark & \checkmark & \checkmark \\
\tauh $\eta$                                &            & \checkmark &                           \\
\Pell $\eta$                                &            & \checkmark &                           \\
leading VBS jet \pt                         & \checkmark & \checkmark & \checkmark                         \\
subleading VBS jet \pt                      & \checkmark & \checkmark & \checkmark                       \\
leading VBS jet mass                        &            & \checkmark & \checkmark \\
subleading VBS jet mass                     &            & \checkmark & \checkmark \\
VBS jet pair $\Delta\phi$                   &            & \checkmark &                           \\
$m_{jj}$                                    & \checkmark & \checkmark &                           \\
$m_{1T}$                                    & \checkmark & \checkmark & \checkmark \\
\Moone                                      & \checkmark & \checkmark & \checkmark \\
$\mT(\tauh, \ptvecmiss)$                    &            &            & \checkmark \\
$\mT(\Pell, \ptvecmiss)$                    & \checkmark & \checkmark & \checkmark \\
$\mT(\Pell+\tauh, \ptvecmiss)$              &            &            & \checkmark \\
\PTrel{\Pell, j_1}                          &                         & \checkmark &                           \\
\PTrel{\Pell, j_2}                          &                         & \checkmark &                           \\
\PTrel{\tauh, j_1}                          &                         & \checkmark &                           \\
\PTrel{\tauh, j_2}                          &                         & \checkmark &                           \\
$\Delta\phi(\Pell,j_1)$                     &                         & \checkmark &                           \\
$\Delta\phi(\Pell,j_2)$                     &                         & \checkmark &                           \\
$\Delta\phi(\tauh,j_1)$                     &                         & \checkmark &                          \\
$\Delta\phi(\tauh,j_2)$                     &                         & \checkmark &                           \\
$p_{\mathrm{T},\, \text{leading }\tauh\text{ track}}/p_{\mathrm{T},\, \tauh}$ & \checkmark & \checkmark & \\
$z_{\text{event}}$                          &                         & \checkmark &                           \\
\hline
\end{tabular}
\label{tab:DNNfeatures}
\end{table}

For the measurement of the SM VBS processes under investigation, the statistical analysis is implemented with an ML fit to extract the signal strength taking the asymptotic limit of Wilks's theorem \cite{Cowan:2010js} to infer the test statistic's dependency on the signal strength.  The ML fit is implemented with the CMS statistical analysis tool \textsc{Combine}~\cite{CAT-23-001}, which is based on the \textsc{RooFit}~\cite{Verkerke:2003ir} and \textsc{RooStats}~\cite{Moneta:2010pm} frameworks. To validate the results obtained by relying on the asymptotic limit, we perform the same measurement by generating pseudo-experiments for the signal and the background, taking into account their statistical fluctuations. The outcomes from the pseudo-experiments are consistent with the ones returned by applying Wilks's theorem, and in the following, the latter are presented. Data yields in both SRs and CRs are incorporated in the likelihood via Poisson probability density functions. The inputs to the fit are the distributions in the DNN output of the data, the signal, and the backgrounds estimated as described above. The distributions in the SRs and CRs are affected by common sources of systematic uncertainty, described in the next section, and thus their expectations are treated as correlated in the fit.

The statistical analysis for the investigation of the EFT contributions in the EW VBS processes of interest is also based on an ML fit, considering that the expected number of events \Nexp inherits the quadratic dependence on the EFT Wilson coefficients from the scattering amplitude $\abs{\mathcal{A}_{\text{BSM}}}$ reported in Eq.~\eqref{eq:absm}. When the contribution of a single EFT dim-6 (with $c_i$ Wilson coefficient) or dim-8 (with $f_\alpha$ Wilson coefficient) operator is considered, and all the others are set to null values, the expected number of events can be written as follows:
\begin{align}
\Nexp^{i} &= N_\mathrm{SM} + \frac{c_{i}}{\Lambda^2}N_\mathrm{Lin}^{i} + \frac{c_{i}^2}{\Lambda^4}N_\text{Quad}^{i}\label{eq:1D6}, \\
\Nexp^{\alpha} &= N_\mathrm{SM} + \frac{f_{\alpha}}{\Lambda^4}N_\mathrm{Lin}^{\alpha} + \frac{f_{\alpha}^2}{\Lambda^8}N_\text{Quad}^{\alpha}\label{eq:1D8}, 
\end{align}
where $N_\mathrm{SM}$ stands for the contribution from the SM processes, $N_\mathrm{Lin}$ for the one from the interference of the considered EFT operator with the SM EW VBS processes, and $N_\text{Quad}$ for the pure term produced by the specific operator. 
In the following, this fit setup will be referred to as a 1D dim-6 or dim-8 EFT study, respectively. When two EFT dim-6 operators with Wilson coefficients $c_{i}, c_{j}$ are considered active, and all the other ones are set as negligible, the same quantity reads:
\begin{equation}
\Nexp^{i,j} = N_\mathrm{SM} + \sum_{k=i,j}\left(\frac{c_{k}}{\Lambda^2}N_\mathrm{Lin}^{k} + \frac{c_{k}^2}{\Lambda^4}N_\text{Quad}^{k}\right) + \frac{c_{i}c_{j}}{\Lambda^4}N_\mathrm{Cross}^{ij}\label{eq:2D66}, \\
\end{equation}
where $N_\mathrm{Cross}$ represents the interference between the two EFT operators under study. This fit setup will be called a two-dimensional (2D) same-dimension EFT study in the rest of this paper. Finally, when considering one EFT dim-6 and one dim-8 operator with Wilson coefficients $c_{i}, f_{\alpha}$ to be active, the expected number of events becomes:
\begin{equation}
\Nexp^{i,\alpha} = N_\mathrm{SM} + \frac{c_{i}}{\Lambda^2}N_\mathrm{Lin}^{i} + \frac{c_{i}^2}{\Lambda^4}N_\text{Quad}^{i} + \frac{f_{\alpha}}{\Lambda^4}N_\mathrm{Lin}^{\alpha} + \frac{f_{\alpha}^2}{\Lambda^8}N_\text{Quad}^{\alpha} \label{eq:2D68}. \\
\end{equation}
This fit setup will be referred to as a 2D different-dimension EFT study. It neglects the contributions due to the possible interference between the dim-6 and the dim-8 operator, for which there are no clear theoretical predictions. The absence of this type of EFT contribution precludes consideration of any linear correlation between the two different classes of operators. Consequently, within this theoretical framework, our physics interpretation is limited to the independent effects arising from the interplay between two operators of different dimensions.
For every EFT statistical study, the signal strength of the SM EW SSWW VBS contribution is fixed to 1.

\section{Systematic uncertainties}\label{sec:sys}
Systematic uncertainties in the signal and background yields are introduced as log-normal or Gaussian nuisance parameters in the ML fit, both for the measurement of the EW VBS SSWW processes and for the EFT investigations.

The uncertainties determined by the CMS luminosity monitoring are partially correlated among the data sets~\cite{lumi16, lumi17, lumi18}, resulting in overall uncertainties of 1.2, 2.3, and 2.5\% for the 2016, 2017, and 2018 integrated luminosities, respectively. This uncertainty affects only the integrated yields, not the shapes of the distributions.

Uncertainties at the matrix-element level, which impact both the normalization and shape of the background and signal processes, are evaluated through separate variations of the renormalization and factorization scales. Specifically, the renormalization and factorization scales are varied independently by a factor of 2 and 0.5. The resulting uncertainties are then combined by taking the envelope of these variations relative to the nominal expectation, without considering the combinations where one scale is varied by a factor of 2.0 and the other one by a factor of 0.5~\cite{Cacciari:2003fi, Catani:2003zt}. These uncertainties are considered uncorrelated across different process categories but are correlated across the data-taking years.

The MC samples are generated using a default PDF set, as mentioned in Section~\ref{sec:samples}, and event weights corresponding to the 100 PDF alternative set members are also stored, evaluated with the \MGvATNLO reweighting technique introduced in Section 3. These are used to evaluate the PDF systematic uncertainties according to the procedure recommended by the PDF4LHC group~\cite{PDFunc}, which is based on the same strategy explained in the previous paragraph. They are correlated among the data-taking years and processes.

Among the possible theoretical uncertainties affecting MC simulations are those related to the QCD-induced parton-shower modeling. These uncertainties are divided into initial-state radiation (ISR) and final-state radiation (FSR) and are estimated by separately varying each of them by the same up and down factor. The resulting shape distortions induced by the alternative scenarios are used to evaluate the magnitude of the associated uncertainties. They are considered correlated among the data-taking years and processes.

Uncertainties in \PQb tagging and mistagging data-to-simulation scale factors (SFs) are applied to reproduce the corresponding efficiencies measured in the data, and implemented in the ML fit as correlated among the data-taking years.

Systematic uncertainties related to the pileup modeling are introduced as a $\pm$4.6\% variation in the total inelastic cross section of 69.2 mb that is used to estimate the data pileup distributions~\cite{CMS:2018mlc}. They are correlated among the data-taking years.

The impact of the uncertainty in the trigger efficiency measurement is estimated by varying the SFs within their uncertainties separately for each data-taking year and final state. This uncertainty is treated as uncorrelated among data-taking years.

In 2016 and 2017, a portion of trigger primitives in the ECAL was associated with the wrong bunch crossing, leading to a trigger mistiming effect and a nonnegligible decrease in the trigger efficiency that is not modeled in the simulated samples~\cite{CMS:2024ppo}. Events have been corrected for this effect with a per-event weight, and the corresponding uncertainties have been propagated throughout the analysis chain. They are correlated among the data-taking years. In the following, we will refer to the effect described in this paragraph as trigger mistiming. 

In simulated events, reconstructed four-momenta of all of the jets are simultaneously varied according to the $\eta$- and \pt-dependent uncertainties in the jet energy scale; they are correlated among data-taking years. These variations are then propagated to the \ptvecmiss. Moreover, to properly evaluate the systematic effect coming from differences in the jet energy resolution between data and simulations, smearing is also applied to the latter by varying the jet resolutions according to their uncertainties; they are uncorrelated among the data-taking years~\cite{CMS:2016lmd}.  Because of the high efficiency of the jet quality requirements, no SFs or associated uncertainties for those are applied.

Systematics related to uncertainties affecting unclustered energy in the calorimeters are included in the fit and correlated among data-taking years.

Systematic uncertainties due to SFs used to match the efficiencies in light lepton reconstruction, identification, isolation, and energy scale and resolution as measured in the MC samples with those observed in data, are evaluated by varying the corresponding event weights by the SF uncertainty. They are uncorrelated over the data-taking years.

For the uncertainty arising from charge sign misreconstruction in the $\Pe\tauh$ channel we assign a uniform uncertainty of 15\% to the distributions of the background processes, consistent with the data-background agreement observed in the OS/DY CR.

Statistical uncertainties related to the \PGt lepton identification SFs and the corrections to their energy scale and resolution~\cite{CMS:2018jrd} induce a systematic effect on the expected signal and background distributions. Their impact is evaluated following a procedure similar to the one applied for light-lepton systematic uncertainties. They are uncorrelated among the data-taking years.

The LO to NLO corrections to the VBS signals come with statistical uncertainties that are propagated to the fit~\cite{CMS-PAPERS-SMP-19-012,Biedermann:2016yds,Biedermann:2017bss}.

For the signal and background processes estimated from simulation, the precision of the modeling is limited by the event count in the MC samples. The corresponding statistical uncertainties are therefore taken as systematic uncertainties applied to each bin of the corresponding distribution, according to the lite Barlow-Beeston method~\cite{BB}.

The estimate of the nonprompt background is affected by the statistical uncertainties of the auxiliary regions used to measure the transfer factors. This uncertainty is propagated by the lite Barlow-Beeston method~\cite{BB}.
We assign a further 30\% normalization uncertainty based on a closure test performed in the nonprompt VR. This uncertainty is treated as uncorrelated among the data-taking years.

In the following, we collectively refer to the statistical uncertainties assigned to the backgrounds, extracted from data CRs or from simulation, as background statistical.  It represents the dominant source of overall uncertainty. The impacts of the systematic uncertainties in the signal strength, as extracted from the ML fit, are summarized in Table~\ref{tab:statunc}. 

\begin{table}[htpb!]
\centering
\topcaption{The impact of each systematic uncertainty, together with the impact of the data statistical uncertainty, on the signal strength \PGm, as extracted from the fit to measure the SM SSWW VBS signal with the DNN output distributions. Upper and lower uncertainties are given for the various sources.}
\begin{tabular}{@{}lll@{}}
\hline
Uncertainty source                                   & $+\Delta \mu$ & $-\Delta \mu$ \\
\hline
Theory (PDF, scales, ISR, FSR)                       & $+0.21$    & $-0.12$   \\
Nonprompt background estimation                      & $+0.062$   & $-0.070$   \\
Jet energy scale, resolution, and identification     & $+0.077$   & $-0.059$   \\
Trigger mistiming                                    & $+0.071$   & $-0.040$   \\
\ttbar normalization                                 & $+0.048$   & $-0.037$   \\
Pileup                                               & $+0.042$   & $-0.010$   \\
Luminosity                                           & $+0.025$   & $-0.026$   \\
LO-to-NLO VBS corrections                            & $+0.037$   & $-0.024$   \\
\tauh energy scale and identification                & $+0.040$   & $-0.010$   \\
\PQb tagging and mistagging                          & $+0.031$   & $-0.008$   \\
Unclustered energy                                   & $+0.007$   & $-0.005$   \\
Charge misidentification                             & $+0.009$   & $-0.009$   \\
Lepton reconstruction, identification, and isolation & $+0.002$   & $-0.003$   \\
MC sample size (bin-by-bin uncertainty)              & $+0.28$    & $-0.28$   \\
\hline
Total systematic                                     & $+0.36$    & $-0.32$   \\
Data sample size                                     & $+0.50$    & $-0.46$   \\
\hline
Total                                                & $+0.62$    & $-0.56$   \\ \hline
\end{tabular}
\label{tab:statunc}
\end{table}

\section{Measurement of the SM SSWW VBS processes}\label{sec:measurement}
\begin{figure}[htpb!]
    \centering
    \includegraphics[width=0.49\linewidth]{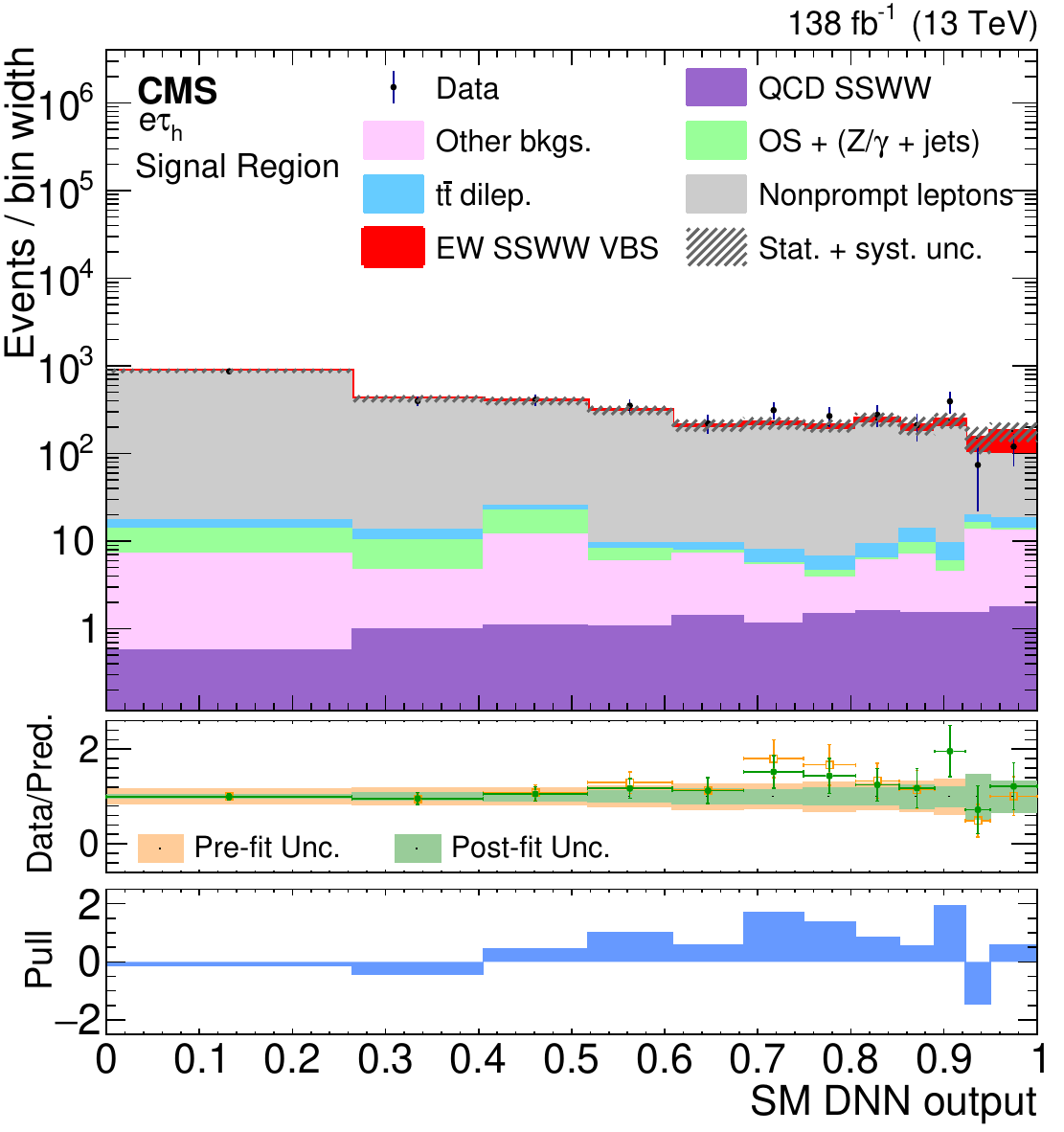}
    \includegraphics[width=0.49\linewidth]{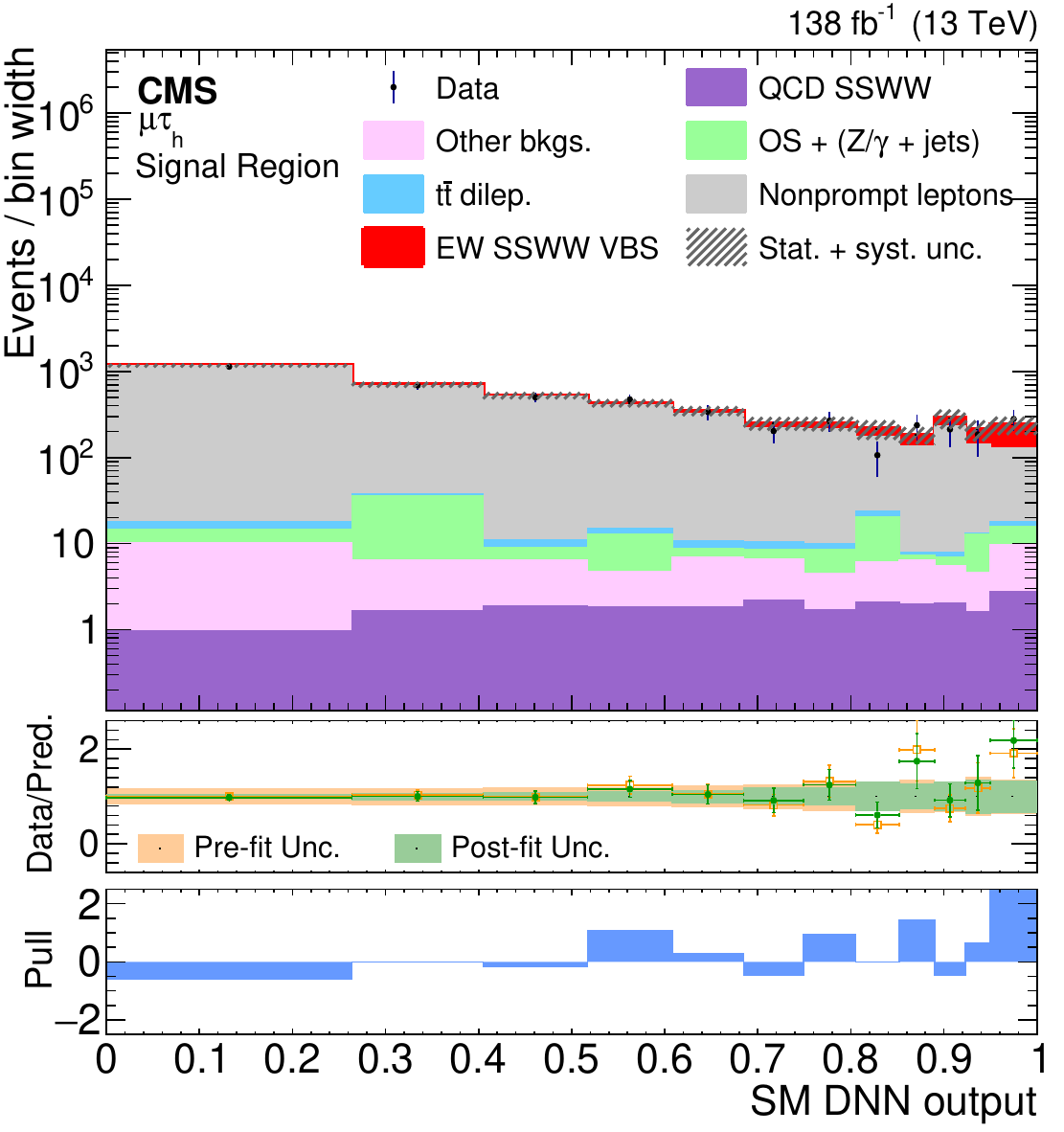}
    \caption{Distribution of the DNN output for the (\cmsLeft) \etauh and (\cmsRight) \mutauh SR. The data points are overlaid on the post-fit background plus signal (stacked histograms). The overflow is included in the last bin. The middle panels show ratios of the data to the pre-fit background prediction and post-fit background yield in yellow and green, respectively. The corresponding colored bands indicate the systematic component of the uncertainty. The lower panels show the distributions of the pulls, defined in the text. The vertical bars represent the statistical uncertainty assigned to the observed number of events for the upper panel, and its propagation to the quantities shown in the middle panel.}
    \label{fig:postfit_DNN_SM_RunII_mutau}
\end{figure}

We extract values of the signal strength and the statistical significance from the fit of the SM SSWW VBS signal, with two separate interpretations. As the primary interpretation, we measure the purely EW signal strength keeping the QCD SSWW production contribution fixed to the SM prediction. In the second interpretation, we measure the signal strength treating as signal the combined EW and QCD SSWW processes, fixing the ratio between the two contributions to the SM value.
For the primary result of the EW signal strength measurement, the post-fit distributions of the output score of the SM DNN (SM DNN output) of the SRs are shown in Fig.~\ref{fig:postfit_DNN_SM_RunII_mutau} for both the electron and muon flavors. 

In Fig.~\ref{fig:postfit_DNN_SM_RunII_mutau}, the data are compared with the background estimated before (pre-fit) and after (post-fit) the simultaneous fit of the SRs and CRs. The pulls shown in the lower panels are defined, for each bin, as
\begin{equation}
    \text{Pull} = \frac{n_\text{data} - n_\text{post-fit}}{\sqrt{\sigma^2_\text{data} - \sigma^2_\text{post-fit}}},
\end{equation}
where $n_\text{data}$ and $n_\text{post-fit}$ are respectively measured event numbers and post-fit background predictions, and $\sigma_\text{data}$ and $\sigma_\text{post-fit}$ are their corresponding uncertainties; when $\sigma_\text{data} < \sigma_\text{post-fit}$, the pull is set to 0. The quadratic difference of the uncertainties appearing in the denominator is taken to account for the correlation between the data and the post-fit prediction.
The observed (expected) EW signal strength is $1.44^{+0.63}_{-0.56}$ ($1.00^{+0.60}_{-0.53}$), corresponding to a signal significance of 2.7 (1.9) standard deviations.
The simultaneous measurement of the EW and QCD-associated SSWW production results in an observed (expected) signal strength equal to $1.43^{+0.60}_{-0.54}$ ($1.00^{+0.57}_{-0.51}$), with a significance of $2.9$ ($2.0$) standard deviations.
The largest contribution to the overall uncertainty is the statistical uncertainty of the data, as reported in Table~\ref{tab:statunc}.

Tabulated results are provided in the HEPData record for this analysis~\cite{hepdata}.
\section{Effective field theory interpretation}\label{sec:eft}

The sensitivity of the measurement to the dim-6 and dim-8 EFT operators, considered one at a time, is estimated from a likelihood scan performed by varying the corresponding Wilson coefficients as they appear in the quadratic parametrization of the signal yields given by Eqs.~\eqref{eq:1D6} and \eqref{eq:1D8}. For this part of the study, we use the distributions of the dim-6 and dim-8 DNN outputs, respectively. No significant deviations from the SM predictions are observed.
The 68 and 95\% confidence intervals on the Wilson coefficients are extracted from the scan and reported in Table~\ref{tab:eftres}.
Although CMS and ATLAS have published dim-8 VBS analyses (as reported in Section~\ref{sec:intro}), and ATLAS has set limits on the dim-6 $c_{W}$ and $c_{HWB}$ Wilson coefficients via the reinterpretation of differential cross section measurements~\cite{ATLAS:2023dkz}, our study for the first time investigates EFT dim-6 effects directly at the reconstruction level.
With this strategy, it is possible to tailor the analysis for maximum sensitivity to EFT-induced deviations in the event kinematics, specifically through DNN models optimized towards EFT effects. In addition, in contrast to reinterpretation approaches, reconstruction-level studies can fully capture correlations among multiple dim-6 operators. As a result, we obtain the first detector-level one- and two-dimensional limits on a broad set of EFT dim-6 contributions in VBS processes.

\begin{table}[htbp!]
\topcaption{Observed and expected 68 and 95\% 1D confidence level (\CL) intervals on the Wilson coefficients associated with the EFT dim-6 and dim-8 operators considered. The results reported here are obtained by fixing the Wilson coefficients other than the one of interest to their SM values in the fit procedure. These limits are reported in units of $\TeV^{-2}$ ($\TeV^{-4}$) for the dim-6 (dim-8) Wilson coefficients.}
\centering
\cmsTable{
\begin{tabular}{cccccc}
\hline
\multicolumn{2}{c}{\multirow{2}{*}{Wilson coefficient}} & \multicolumn{2}{c}{68\% \CL interval(s)}       & \multicolumn{2}{c}{95\% \CL interval}                          \\ 
                                    &                    & Observed                                  & Expected                                  & Observed              & Expected              \\
\hline
\multirow{10}{*}{dim-6}             & $c_{ll}^{(1)}/\Lambda^{2}$     & $[-11.6, 0.045]$                          & $[-12.9, -8.03] \cup [-2.95, 1.91]$       & $[-13.5, 2.11]$       & $[-14.6, 3.53]$       \\
                                    & $c^{(1)}_{qq}/\Lambda^{2}$     & $[-0.341, 0.416]$                         & $[-0.501, 0.576]$                         & $[-0.605, 0.681]$     & $[-0.742, 0.818]$     \\
                                    & $c_{W}/\Lambda^{2}$            & $[-0.513, 0.481]$                         & $[-0.681, 0.669]$                         & $[-0.842, 0.818]$     & $[-0.987, 0.974]$     \\
                                    & $c_{HW}/\Lambda^{2}$           & $[-5.48, 4.31]$                           & $[-7.00, 6.09]$                           & $[-8.68, 7.60]$       & $[-9.99, 9.05]$       \\
                                    & $c_{HWB}/\Lambda^{2}$          & $[-30.7, 89.2]$                           & $[-41.7, 69.6]$                           & $[-49.7, 110]$       & $[-66.6, 96.4]$       \\
                                    & $c_{H\Box}/\Lambda^{2}$        & $[-12.0, 14.0]$                           & $[-16.6, 18.1]$                           & $[-20.9, 22.7]$       & $[-24.7, 26.3]$       \\
                                    & $c_{HD}/\Lambda^{2}$           & $[-15.3, 31.5]$                           & $[-24.6, 34.7]$                           & $[-31.4, 45.5]$       & $[-38.2, 48.8]$       \\
                                    & $c_{Hl}^{(1)}/\Lambda^{2}$     & $[-38.2, 39.5]$                           & $[-28.8, 29.9]$                           & $[-69.3, 68.3]$       & $[-49.4, 49.7]$       \\
                                    & $c_{Hl}^{(3)}/\Lambda^{2}$     & $[-0.045, 8.58]$                           & $[-1.43, 2.23] \cup [5.88, 9.54]$         & $[-1.59, 9.94]$       & $[-2.64, 10.8]$       \\
                                    & $c_{Hq}^{(1)}/\Lambda^{2}$     & $[-3.27, 3.44]$                           & $[-4.53, 4.42]$                           & $[-5.55, 5.60]$       & $[-6.56, 6.44]$       \\
\multirow{10}{*}{dim-8}             & $f_{T0}/\Lambda^{4}$           & $[-0.774, 0.842]$                         & $[-1.02, 1.08]$                           & $[-1.32, 1.38]$     & $[-1.52, 1.58]$     \\
                                    & $f_{T1}/\Lambda^{4}$           & $[-0.319, 0.381]$                         & $[-0.426, 0.480]$                         & $[-0.552, 0.613]$     & $[-0.640, 0.695]$     \\
                                    & $f_{T2}/\Lambda^{4}$           & $[-0.851, 1.12]$                         & $[-1.15, 1.37]$                           & $[-1.51, 1.76]$       & $[-1.75, 1.98]$      \\
                                    & $f_{M0}/\Lambda^{4}$           & $[-8.07, 7.70]$                          & $[-9.89, 9.74]$                           & $[-13.1, 12.8]$       & $[-14.6, 14.5]$      \\
                                    & $f_{M1}/\Lambda^{4}$           & $[-9.54, 11.15]$                         & $[-12.5, 13.3]$                           & $[-16.4, 17.7]$       & $[-18.7, 19.6]$      \\
                                    & $f_{M7}/\Lambda^{4}$           & $[-17.6, 15.3]$                          & $[-20.3, 19.2]$                           & $[-27.6, 25.8]$       & $[-29.9, 28.8]$      \\
                                    & $f_{S0}/\Lambda^{4}$           & $[-9.60, 9.82]$                           & $[-11.6, 12.0]$                           & $[-15.9, 16.1]$       & $[-17.4, 17.9]$       \\
                                    & $f_{S1}/\Lambda^{4}$           & $[-40.9, 41.3]$                          & $[-37.4, 38.8]$                           & $[-60.9, 61.8]$       & $[-57.2, 58.6]$       \\
                                    & $f_{S2}/\Lambda^{4}$           & $[-40.9, 41.3]$                           & $[-37.4, 38.8]$                           & $[-60.9, 61.8]$       & $[-57.2, 58.6]$    \\
\hline
\end{tabular}
}
\label{tab:eftres}
\end{table}

\begin{figure}[htpb!]
    \centering
    \includegraphics[width=0.32\linewidth]{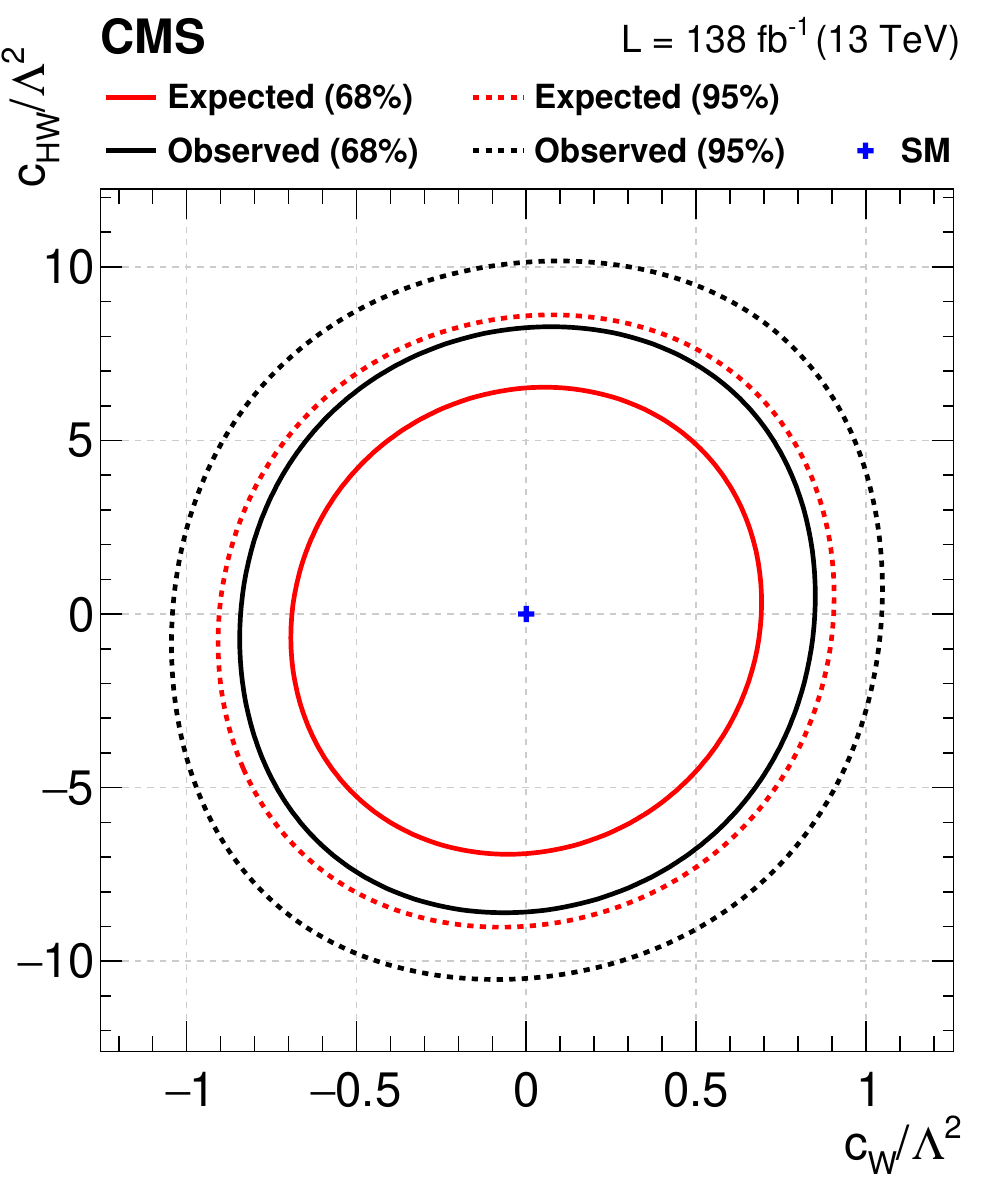}
    \includegraphics[width=0.32\linewidth]{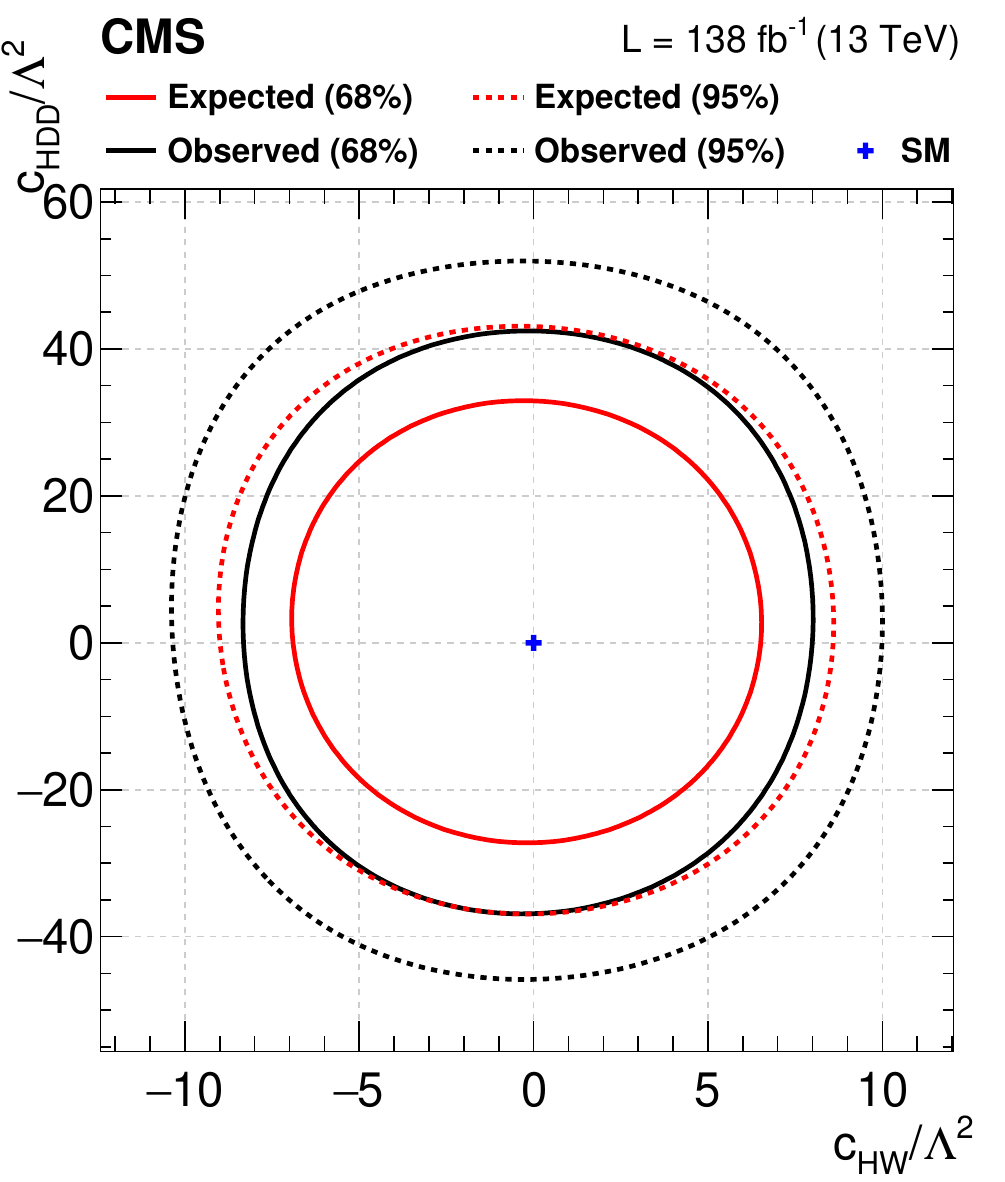}
    \includegraphics[width=0.32\linewidth]{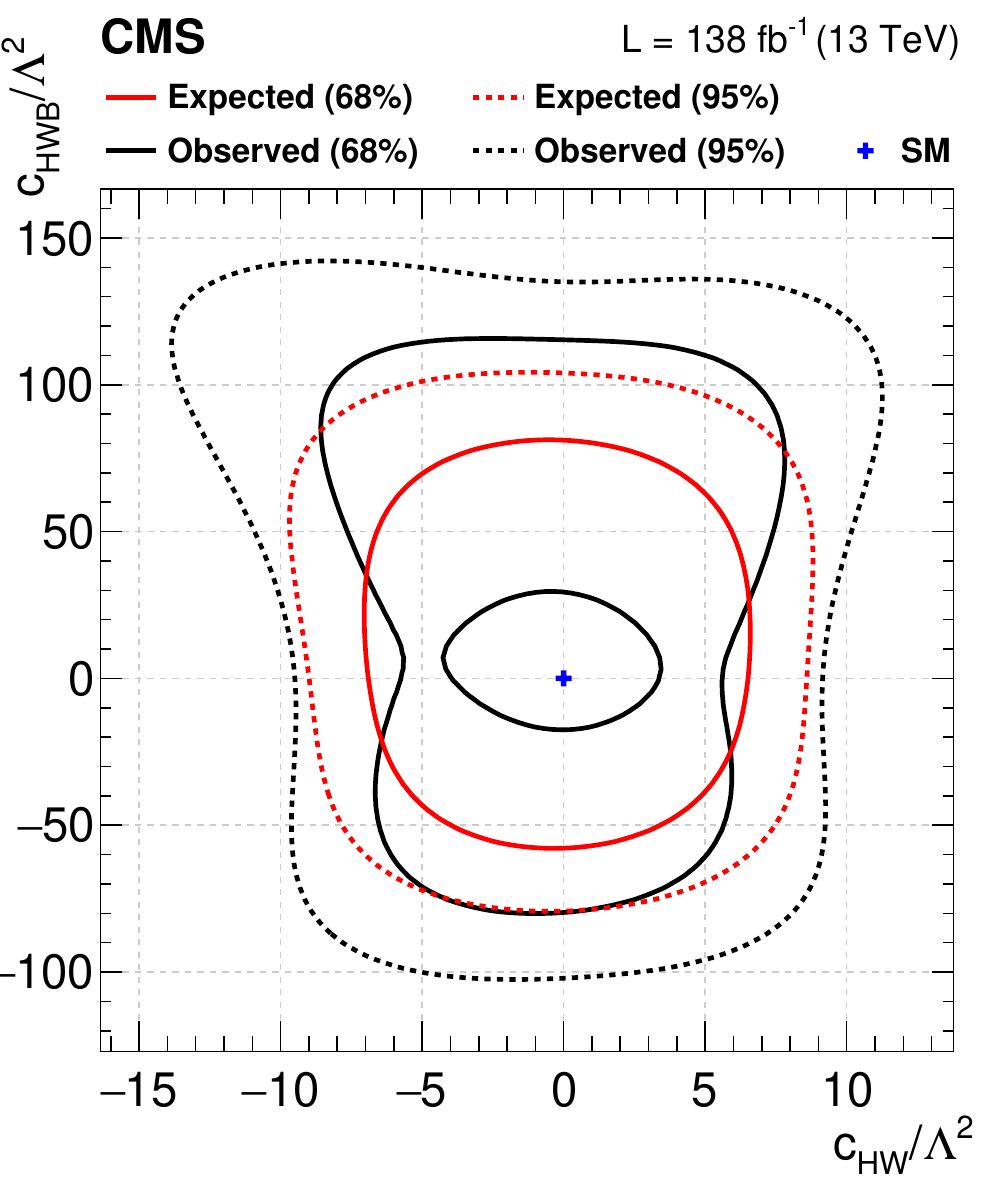}  \\
    \includegraphics[width=0.32\linewidth]{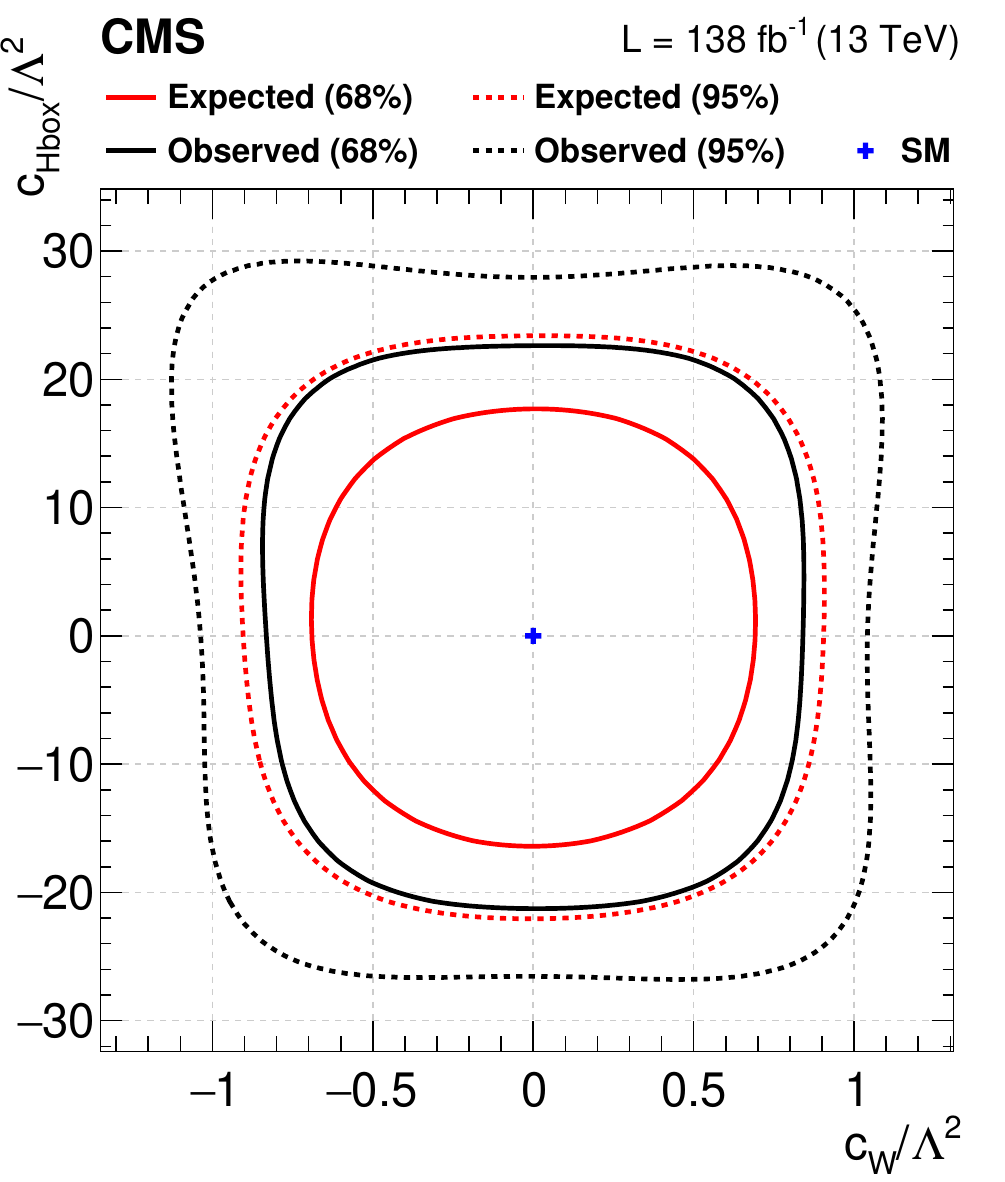}
    \includegraphics[width=0.32\linewidth]{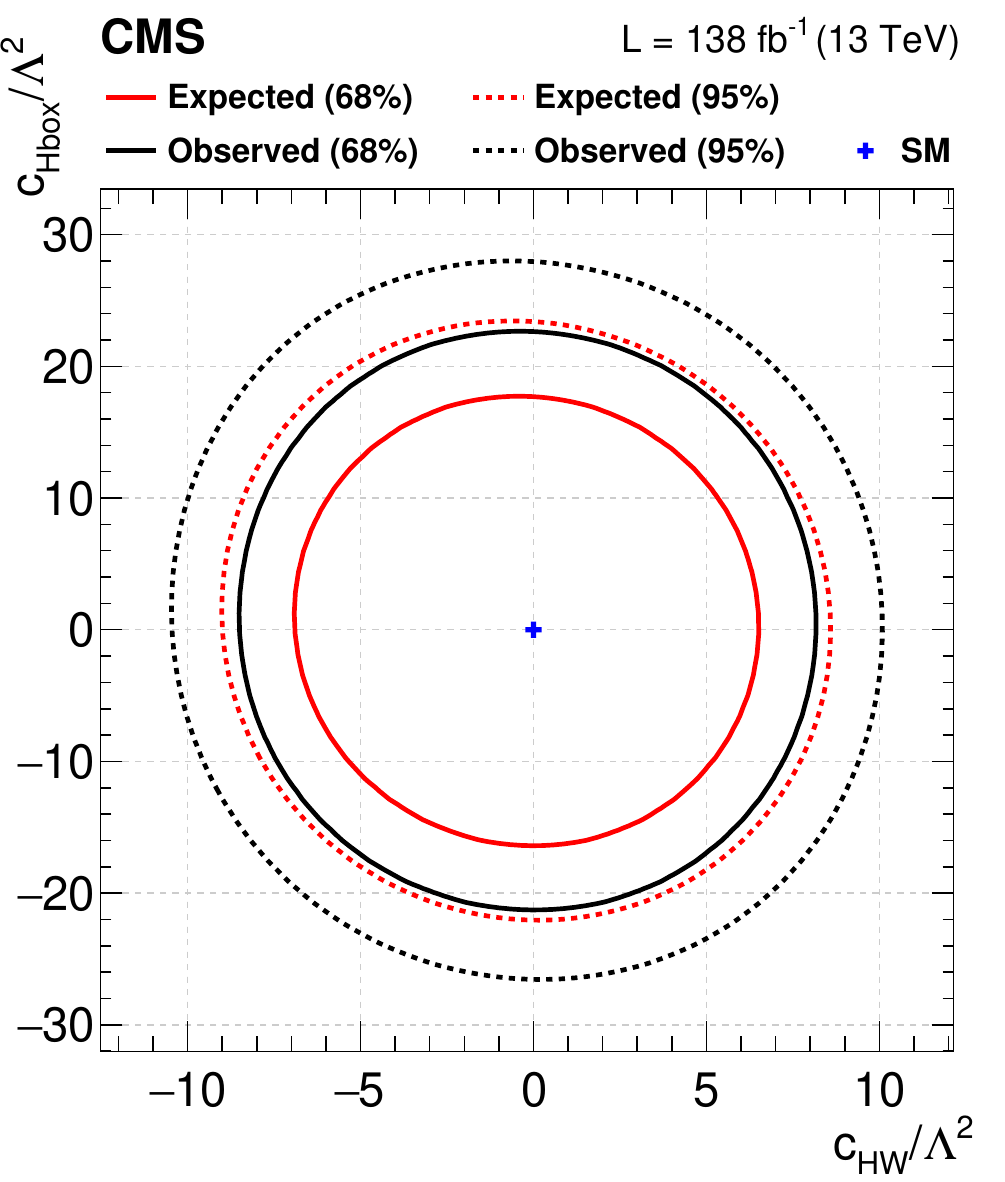}
    \includegraphics[width=0.32\linewidth]{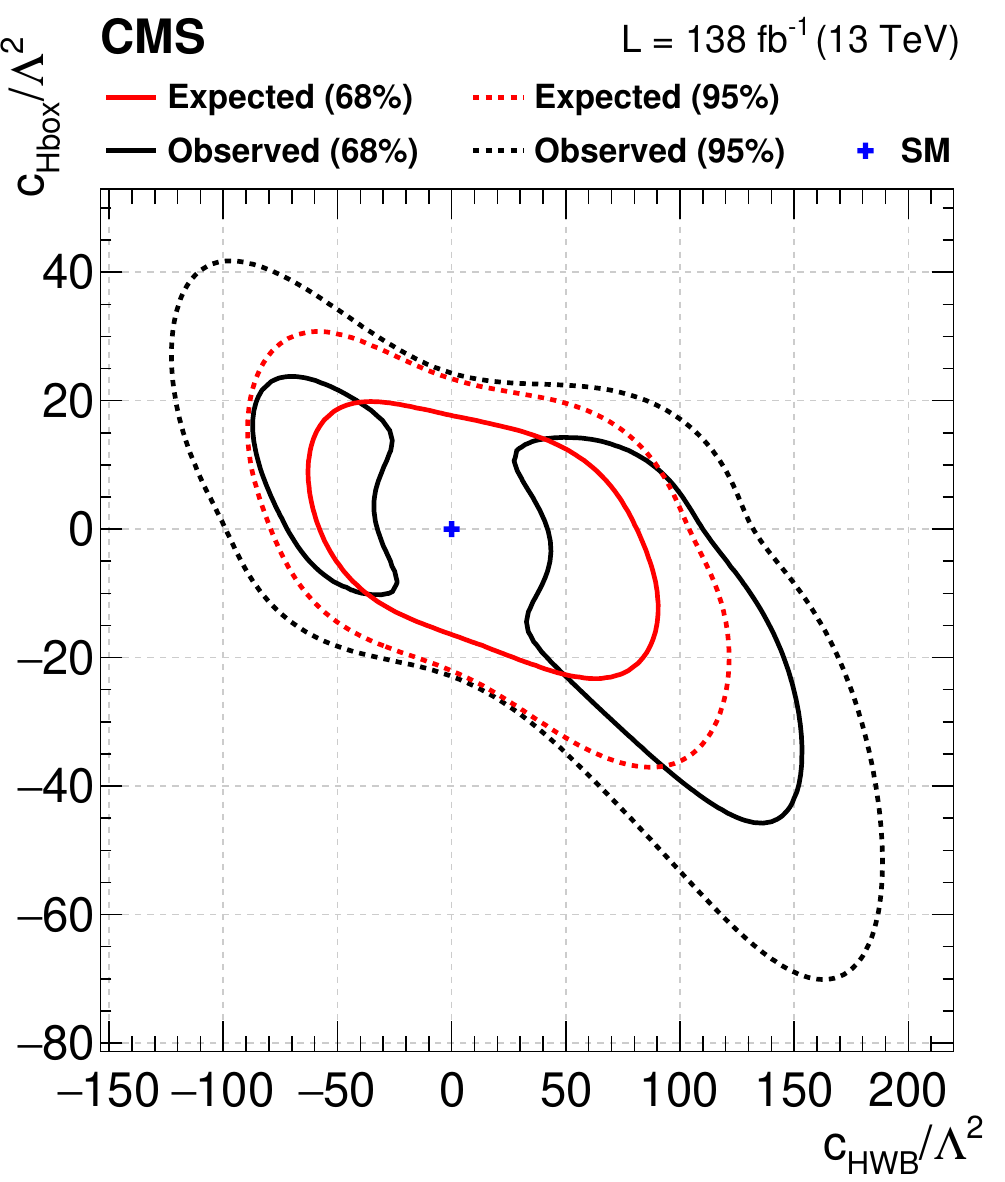} \\
    \includegraphics[width=0.32\linewidth]{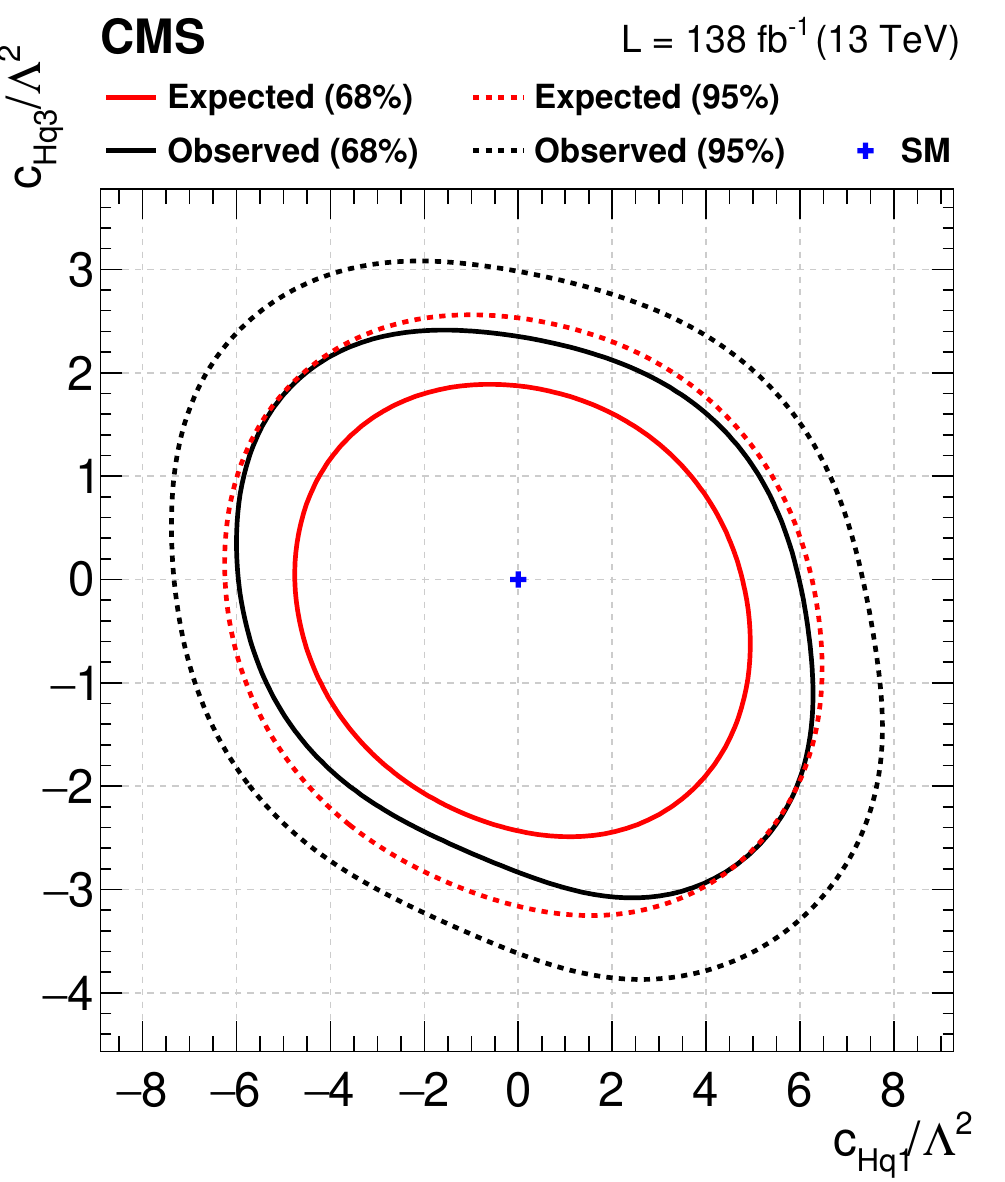} \\
    \caption{Observed (black) and expected (red) 68 (solid) and 95\% (dashed) \CL contours for $-2\ln\Delta\mathcal{L}$ as functions of the reported dim-6 bosonic (upper two rows) and mixed (lower row) Wilson coefficient pairs. When there are two contours for the same \CL value, the constrained set of Wilson coefficient values is represented by the area between the two of them if they are concentric, otherwise it consists of the internal areas of the contours.}
    \label{tab:2Dbosonmixed}
\end{figure}

\begin{figure}[htpb!]
    \centering
    \includegraphics[width=0.32\linewidth]{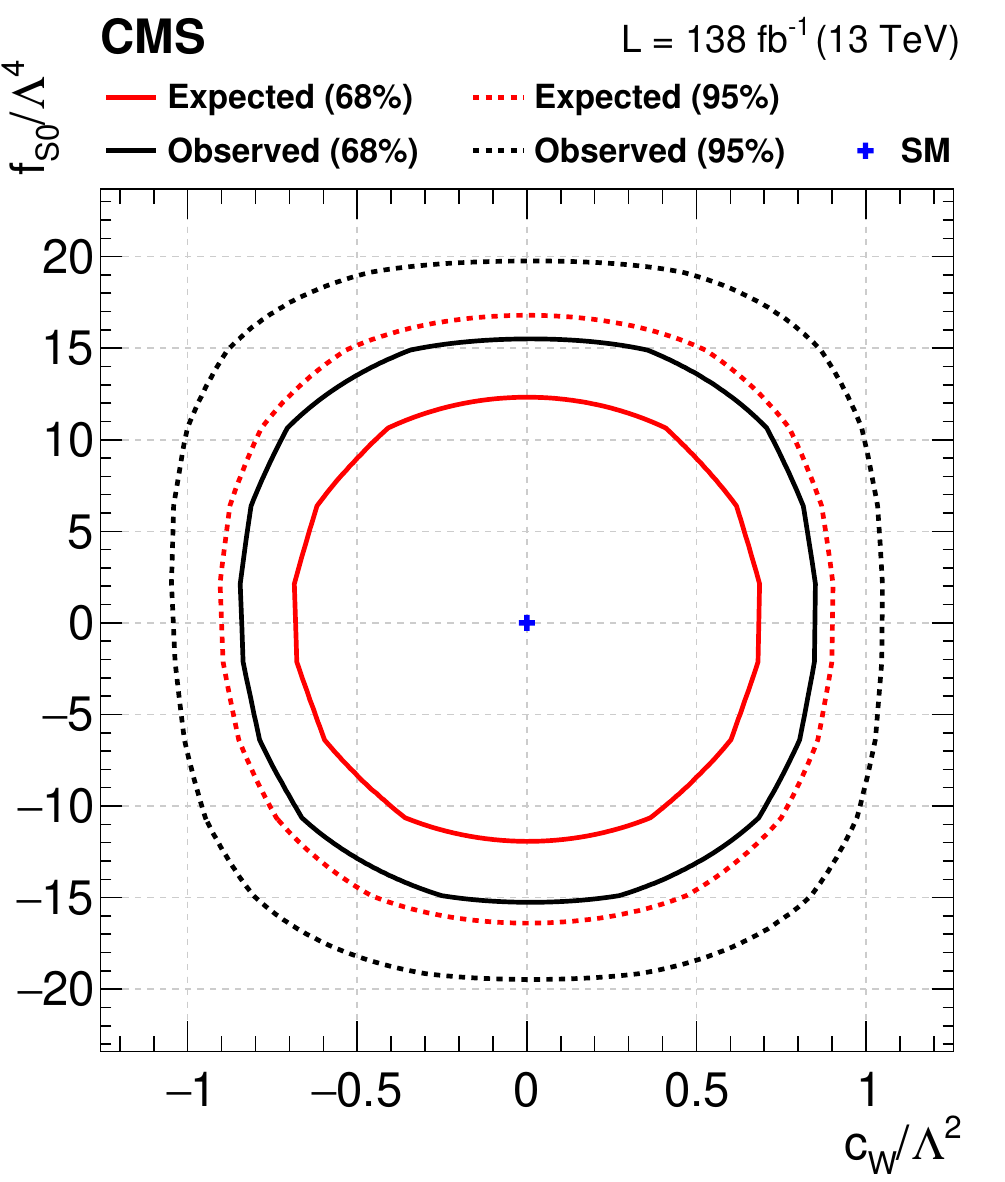}
    \includegraphics[width=0.32\linewidth]{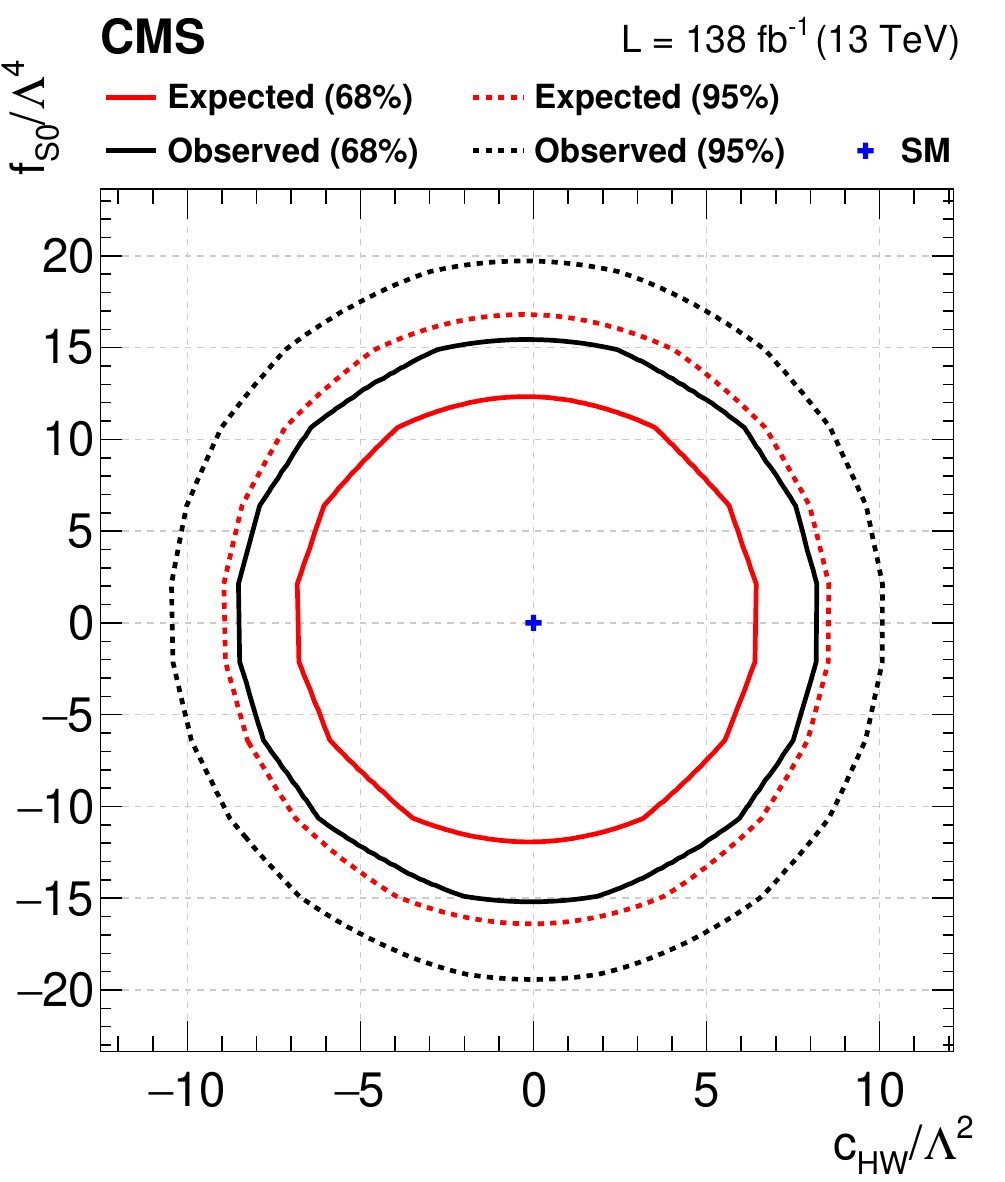}
    \includegraphics[width=0.32\linewidth]{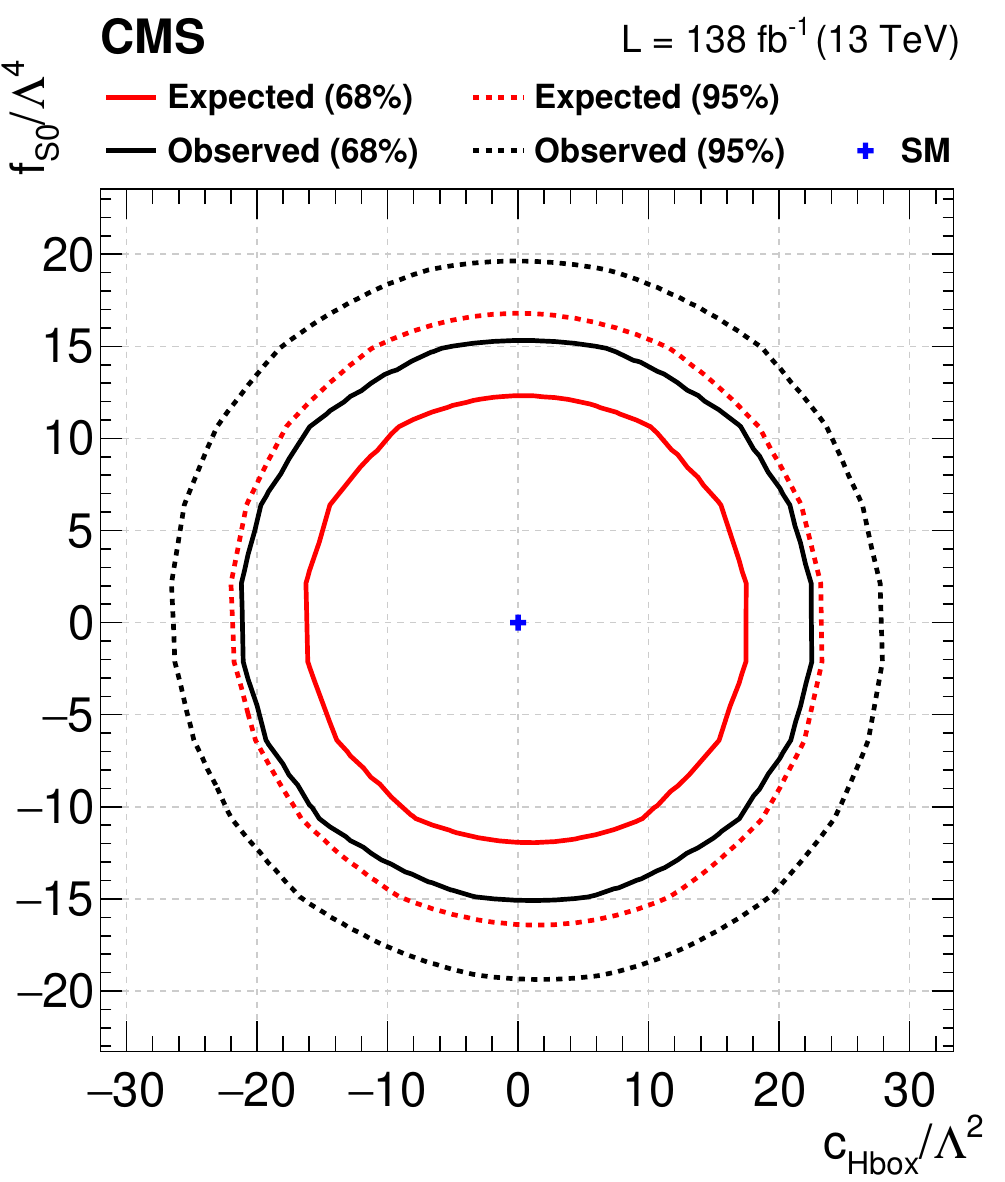} \\
    \includegraphics[width=0.32\linewidth]{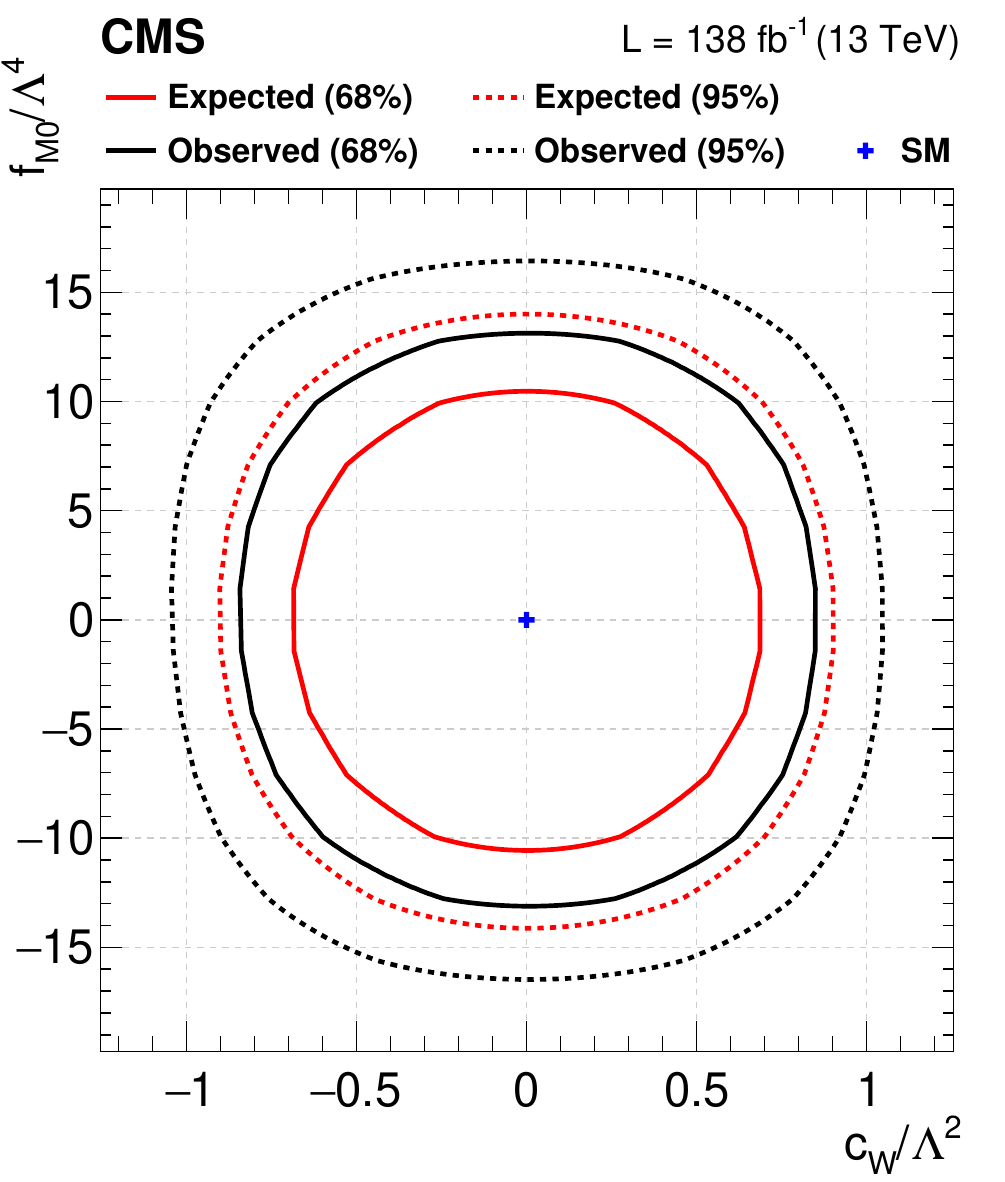}
    \includegraphics[width=0.32\linewidth]{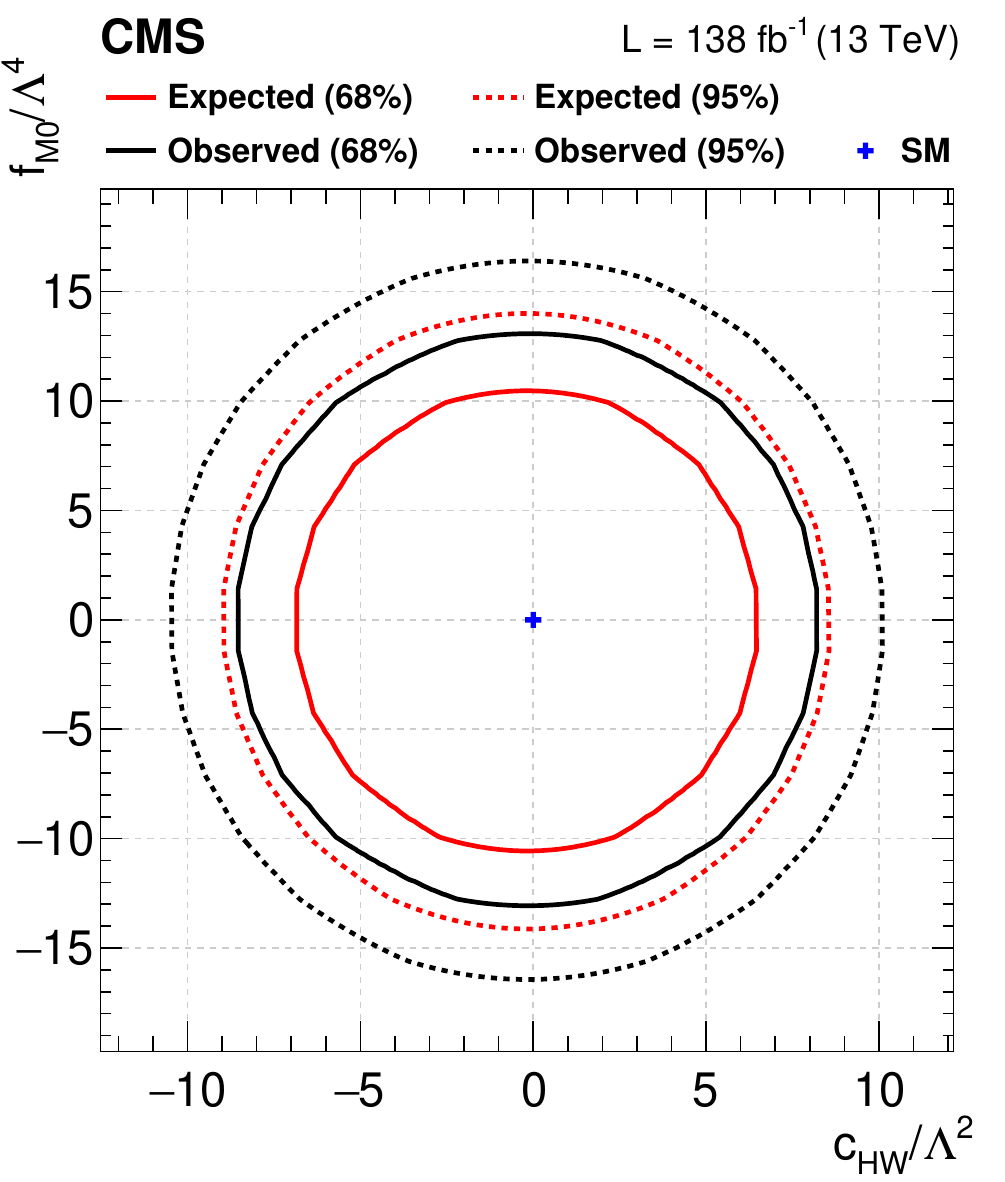}
    \includegraphics[width=0.32\linewidth]{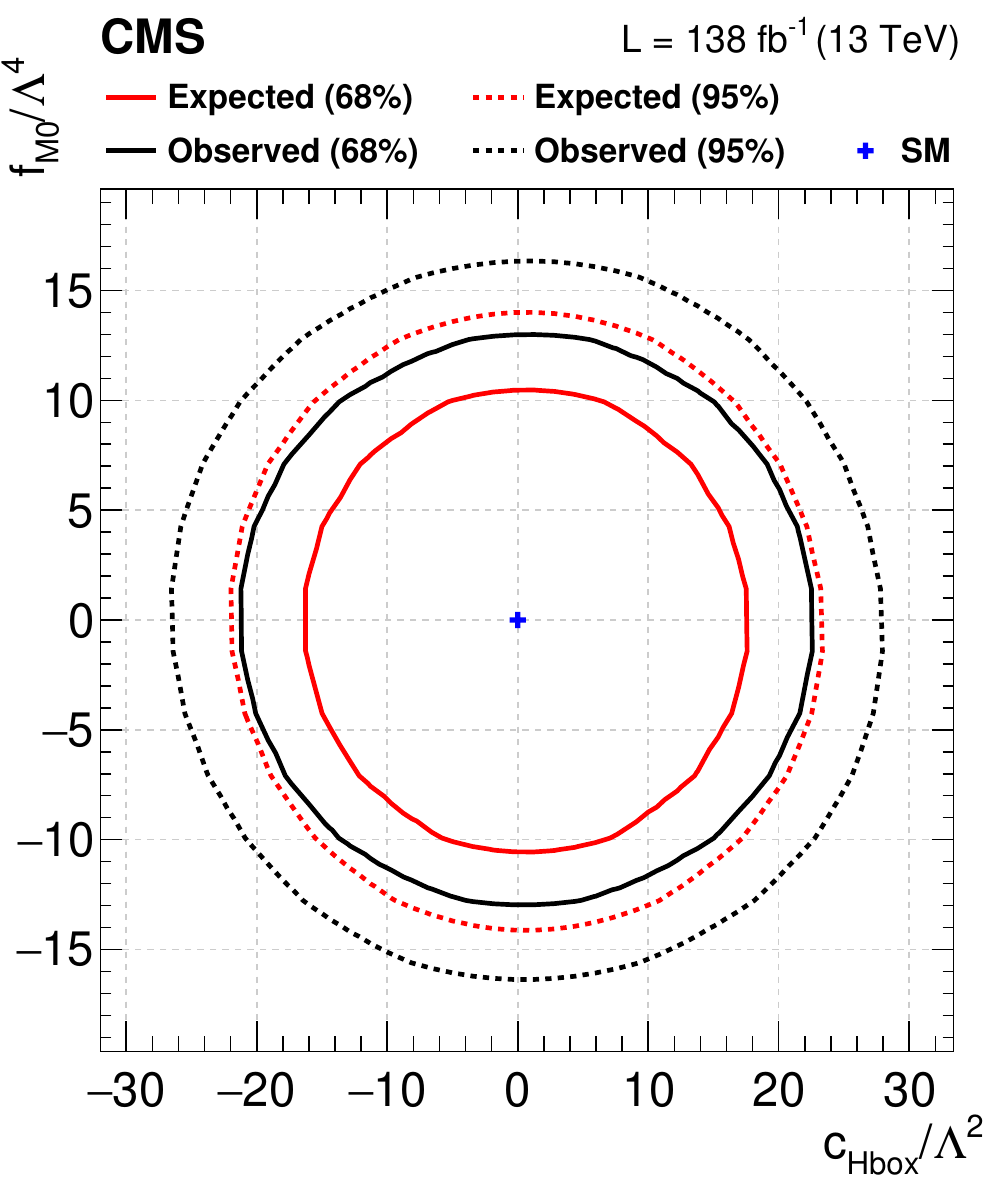} \\
    \includegraphics[width=0.32\linewidth]{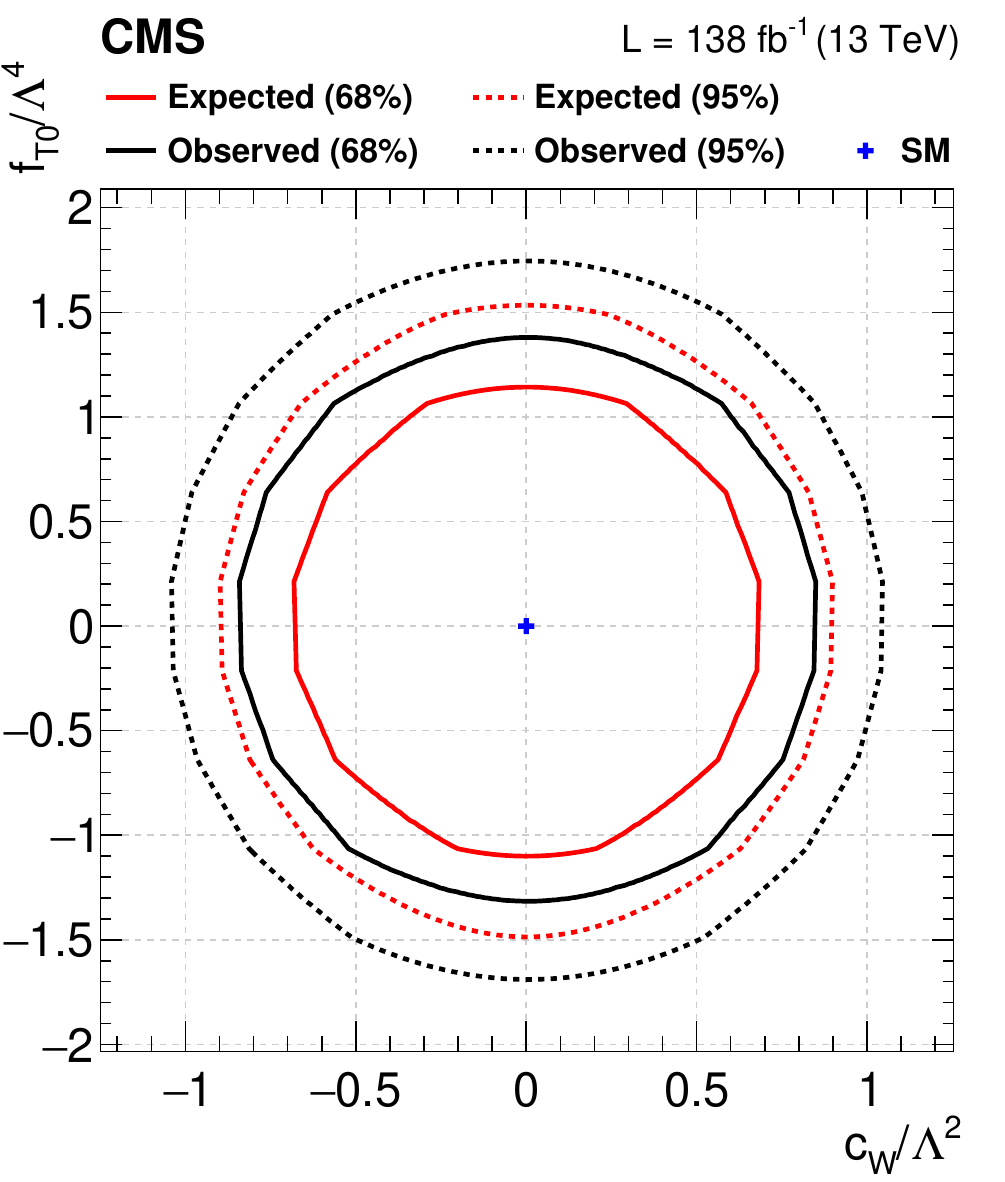}
    \includegraphics[width=0.32\linewidth]{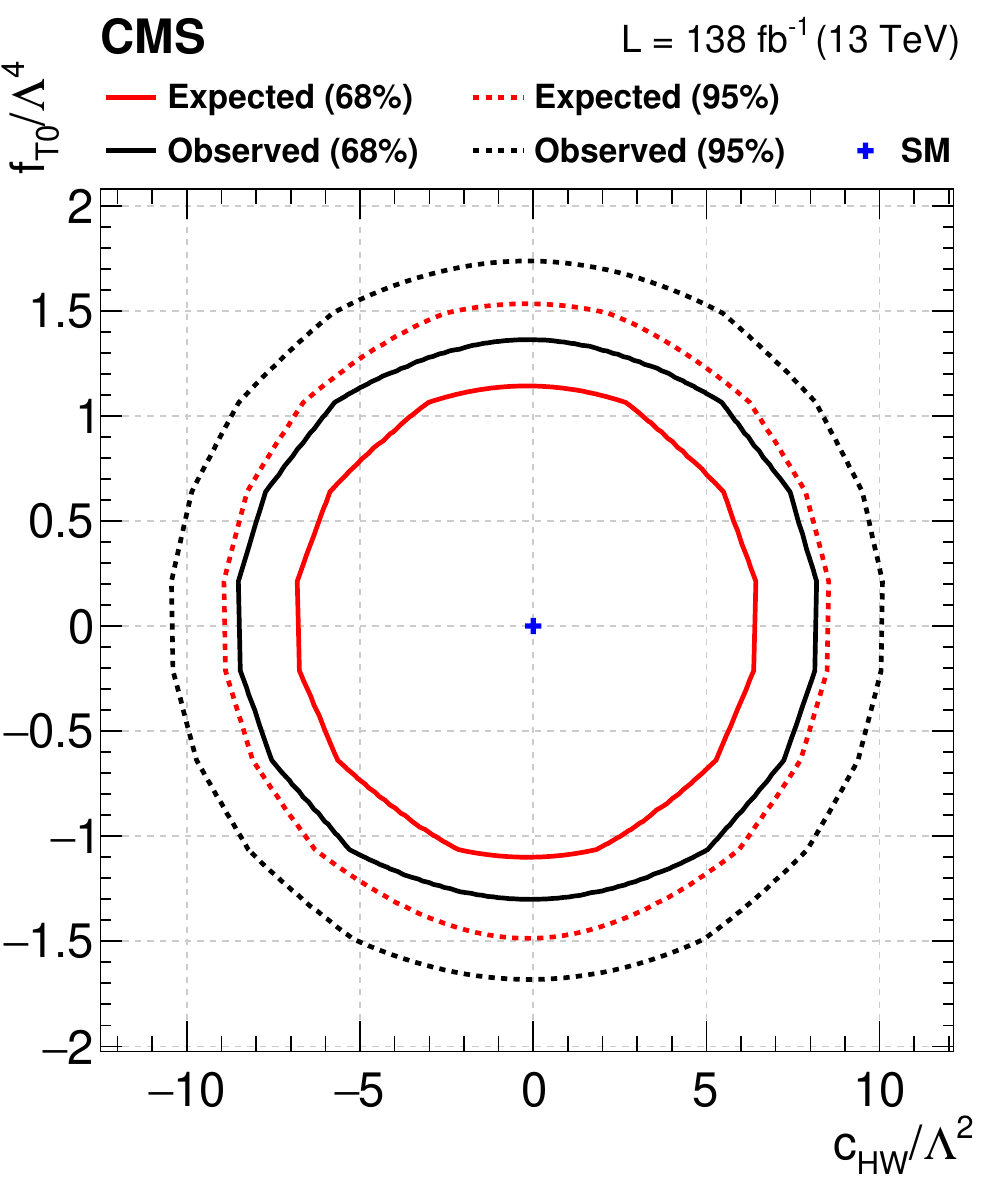}
    \includegraphics[width=0.32\linewidth]{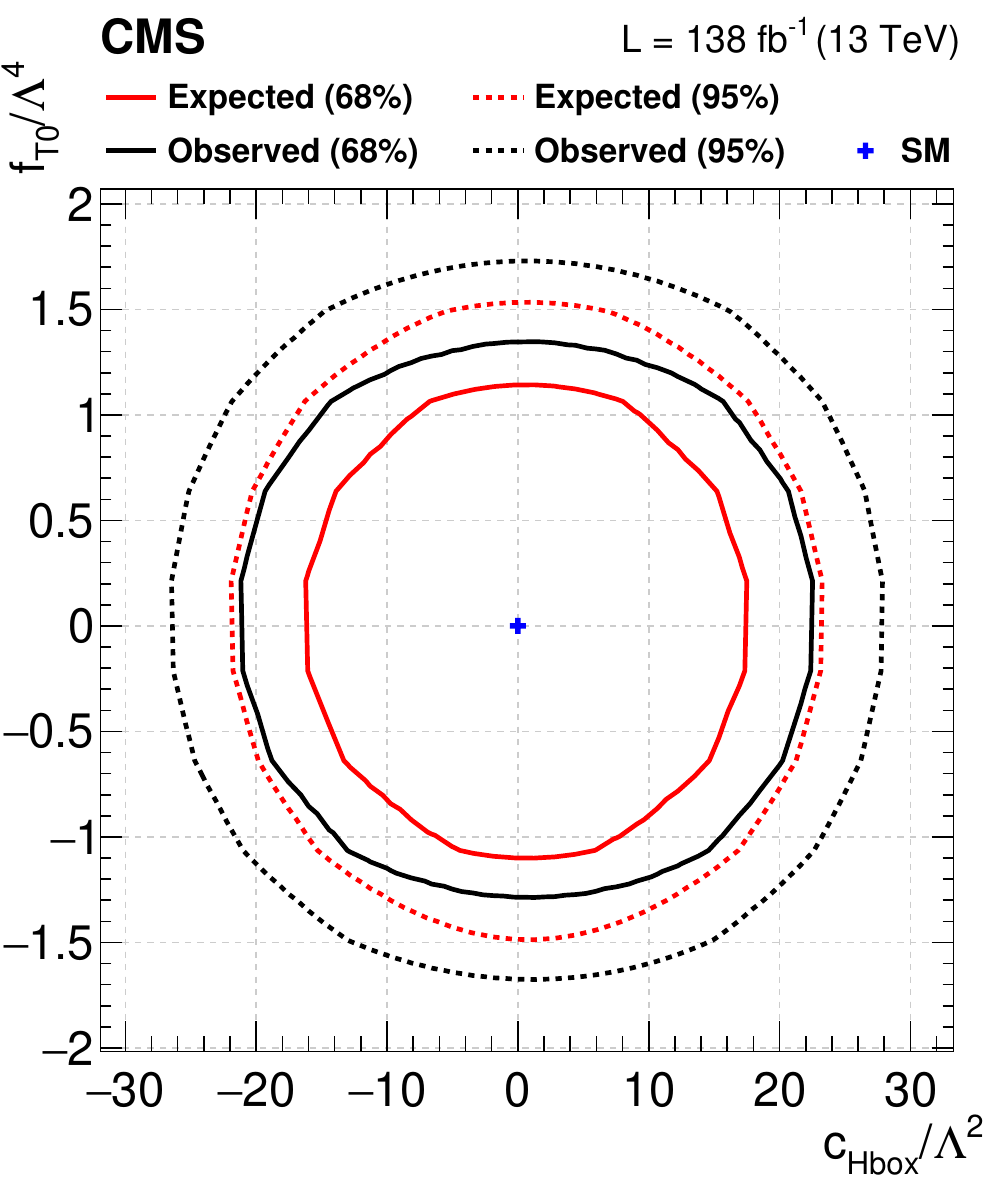} \\
    \caption{Observed (black) and expected (red) 68 (solid) and 95\% (dashed) \CL contours for $-2\ln\Delta\mathcal{L}$ as functions of the reported (dim-6, dim-8) Wilson coefficient pairs.}
    \label{tab:2Dboson68}
\end{figure}

In addition, a 2D likelihood scan is performed over pairs of Wilson coefficients as they appear in Eqs.~\eqref{eq:2D66} and~\eqref{eq:2D68}, exploiting the distributions of \Moone, which we find to be the most sensitive variable to EFT effects among the kinematic quantities considered. This choice ensures the stability of the 2D fits and at the same time enhances the physical interpretability of the results in the broader context of EFT reinterpretation. For the same-dimension fits, we consider pairs of EFT operators that both modify either the $\PW\PW \to \PW\PW$ amplitude or the \PW pairing with fermions. For the different-dimension fits, we consider pairs that modify the $\PW\PW \to \PW\PW$ amplitude. 
The operators in these pairs typically impact the scattering amplitude with similar magnitudes, when the dim-6 (dim-8) coefficients are set to $\TeV^{-2}$ ($\TeV^{-4}$).
The results are reported in Figs.~\ref{tab:2Dbosonmixed} and \ref{tab:2Dboson68}. 

For the same-dimension pairs shown in Fig.~\ref{tab:2Dbosonmixed}, combining the effects of two operators leads to a broadening of the 68 and 95\% \CL intervals extracted with the corresponding one-dimensional fit to the \Moone distributions, an effect that is slightly more pronounced for the operators to which this analysis is less sensitive. On average, this effect results in a 10\% difference for the observed limits and a 13\% difference for the expected limits.
In general, the contours exhibit an elliptical shape, with correlation effects driven by the interference between the operators in each pair, leading to either negative or positive linear correlations that tilt the axes of the ellipses. Additionally, for some operators, the interference with the SM contribution is more significant than the quadratic term, resulting in a double-minima structure in the two-dimensional $-2\ln\Delta\mathcal{L}$ profiles in the direction of the Wilson coefficient associated with the operator introducing such effect. For the observed contours, these effects are further enhanced by data fluctuations that the EFT operators can accommodate.
These effects, either individually or in combination, are visible in the upper right, middle right, and lower panels of Fig.\ref{tab:2Dbosonmixed}, where operators $\mathcal{Q}_{HWB}$, $\mathcal{Q}_{H\Box}, \mathcal{Q}_{Hl}^{(3)}, \mathcal{Q}_{Hq}^{(1)}$ are involved. Notably, they are most pronounced for the $\mathcal{Q}_{HWB}$, $\mathcal{Q}_{H\Box}$ pair (middle right of Fig.\ref{tab:2Dbosonmixed}), where significant mutual interference and relatively more important linear terms (compared to the quadratic contributions at unity Wilson coefficient values) are observed.

For the different-dimension pairs, the results indicate that adding the linear and quadratic dim-8 contributions to those of a given dim-6 operator leads to a broadening of the one-dimensional 68\% and 95\% confidence intervals for the dim-6 operators extracted from the same \Moone distributions, regardless of the specific dim-8 operator considered. This broadening is quantified by comparing the 1D intervals obtained from individual fits with the intervals derived from the 2D limits for one coefficient when the other is fixed at its null value.
On average, this effect on the dim-6 Wilson coefficients results in a 10\% difference in the observed limits and a 12\% difference in the expected limits. A similar broadening is observed for the dim-8 Wilson coefficients in the pairs displayed in Fig. 6, with average differences of 10\% for the observed limits and 13\% for the expected limits.
Moreover, by setting $c/\Lambda^2=1$ for the dim-6 operator and $f/\Lambda^4=1$ for the dim-8 operator, we can directly compare the effects of the dim-6 and dim-8 operators in each pair and relate them to the shape of the corresponding 2D contours. When the contributions of these operators are comparable, the 68\% and 95\% confidence contours in the 2D fit appear more circular; conversely, if the effect of the dim-8 operator is negligible relative to that of the dim-6 operator, the contours tend to be more rectangular. For example, in the bottom left of Fig. 6 the 2D fit for operators $\mathcal{Q}_{W}$ and $\mathcal{O}_{T0}$ shows circular contours when the effects are similar, whereas the plot on top left of Fig. 6 illustrates that when the dim-8 contribution of the $\mathcal{O}_{S0}$ operator is negligible compared to the dim-6 contribution of the operator $\mathcal{Q}_{W}$, the contours become rectangular.
In addition, the contour does not show evidence of linear correlations, as the possible interference between the dim-6 and the dim-8 operators in the pair is neglected in this study.

These considerations lead to the conclusion that, within the EFT framework, the one-dimensional constraints on a Wilson coefficient associated with a given operator could be biased by neglect of the contributions of other operators arising in the same physical process and at the same or different power of $\Lambda_{\text{BSM}}$,  with the same or different dimension.
This bias is not an experimental artifact but stems from the fundamental structure of the SMEFT model. Contributions from dim-6 and dim-8 operators can interfere nontrivially and can be visible at the reconstruction level. This finding aligns with the EFT theory community’s recommendation to perform studies that simultaneously investigate dim-6 and dim-8 contributions and to encompass as wide a range as possible of operator classes and processes~\cite{Hays:2018zze, Dawson:2024ozw, Corbett:2023qtg, CMS:2025ugn, ATLAS:2022xyx}.

It is worth noting that the 2D EFT fits with two dim-6 operators represent the first results of this type for an analysis investigating EW VBS processes, and the 2D EFT fits with one dim-6 and one dim-8 operator represent the first results ever for combinations of different-dimension EFT operators in the same physics processes.

\section{Summary}\label{sec:summary}
Electroweak (EW) production of a same-sign \PW boson pair in association with two jets, with a hadronically decaying \PGt lepton in the final state, is investigated for the first time, together with an interpretation of possible deviations from the standard model expectations in terms of effective field theory (EFT) operators of dimension 6 and 8.
The analysis is performed with a sample of proton-proton collisions at $\sqrt{s} = 13\TeV$ recorded by the CMS experiment at the CERN LHC in 2016--2018, corresponding to an integrated luminosity of 138\fbinv.
Events are selected with the requirement of one \PGt lepton together with one light lepton (\Pe or \PGm) of the same sign, missing transverse momentum, and two jets with large pseudorapidity separation and large dijet invariant mass.  Deep neural network algorithms are employed to discriminate different types of signal events from the main backgrounds, significantly boosting the sensitivity of the search.

The amplitude for same-sign \WW production includes terms that account for strong interactions between partons with \PW boson radiation.  A small fraction of these QCD-mediated events falls within the acceptance of the search.  The measured cross section for EW same-sign \WW scattering, extracted with the QCD-mediated amplitudes fixed to the standard model (SM) expectations,
is $1.44^{+0.63}_{-0.56}$ times the SM prediction. The observed (expected) significance of the EW signal is 2.7 (1.9) standard deviations. A measurement of the combined EW and residual QCD-mediated contributions yields an observed (expected) significance of 2.9 (2.0) standard deviations.

Also presented are the limits on dimension-6 EFT operator contributions, including both one operator and two operators active at the same time, extracted with a reconstruction-level strategy for the first time in vector boson scattering.
This is the first study of the combined effects of EFT operators with different dimensions, showing that focusing on one dimensionality can lead to an overestimate of the sensitivity to the corresponding EFT operator class, and that the contributions of terms combining operators with different dimensions should not be neglected, in accordance with the latest recommendations from the EFT theory community.

\begin{acknowledgments}
    We congratulate our colleagues in the CERN accelerator departments for the excellent performance of the LHC and thank the technical and administrative staffs at CERN and at other CMS institutes for their contributions to the success of the CMS effort. In addition, we gratefully acknowledge the computing centers and personnel of the Worldwide LHC Computing Grid and other centers for delivering so effectively the computing infrastructure essential to our analyses. Finally, we acknowledge the enduring support for the construction and operation of the LHC, the CMS detector, and the supporting computing infrastructure provided by the following funding agencies: SC (Armenia), BMBWF and FWF (Austria); FNRS and FWO (Belgium); CNPq, CAPES, FAPERJ, FAPERGS, and FAPESP (Brazil); MES and BNSF (Bulgaria); CERN; CAS, MoST, and NSFC (China); MINCIENCIAS (Colombia); MSES and CSF (Croatia); RIF (Cyprus); SENESCYT (Ecuador); ERC PRG, RVTT3 and MoER TK202 (Estonia); Academy of Finland, MEC, and HIP (Finland); CEA and CNRS/IN2P3 (France); SRNSF (Georgia); BMBF, DFG, and HGF (Germany); GSRI (Greece); NKFIH (Hungary); DAE and DST (India); IPM (Iran); SFI (Ireland); INFN (Italy); MSIP and NRF (Republic of Korea); MES (Latvia); LMTLT (Lithuania); MOE and UM (Malaysia); BUAP, CINVESTAV, CONACYT, LNS, SEP, and UASLP-FAI (Mexico); MOS (Montenegro); MBIE (New Zealand); PAEC (Pakistan); MES and NSC (Poland); FCT (Portugal); MESTD (Serbia); MCIN/AEI and PCTI (Spain); MOSTR (Sri Lanka); Swiss Funding Agencies (Switzerland); MST (Taipei); MHESI and NSTDA (Thailand); TUBITAK and TENMAK (Turkey); NASU (Ukraine); STFC (United Kingdom); DOE and NSF (USA).
     
    \hyphenation{Rachada-pisek} Individuals have received support from the Marie-Curie program and the European Research Council and Horizon 2020 Grant, contract Nos.\ 675440, 724704, 752730, 758316, 765710, 824093, 101115353, 101002207, and COST Action CA16108 (European Union); the Leventis Foundation; the Alfred P.\ Sloan Foundation; the Alexander von Humboldt Foundation; the Science Committee, project no. 22rl-037 (Armenia); the Belgian Federal Science Policy Office; the Fonds pour la Formation \`a la Recherche dans l'Industrie et dans l'Agriculture (FRIA-Belgium); the F.R.S.-FNRS and FWO (Belgium) under the ``Excellence of Science -- EOS" -- be.h project n.\ 30820817; the Beijing Municipal Science \& Technology Commission, No. Z191100007219010 and Fundamental Research Funds for the Central Universities (China); the Ministry of Education, Youth and Sports (MEYS) of the Czech Republic; the Shota Rustaveli National Science Foundation, grant FR-22-985 (Georgia); the Deutsche Forschungsgemeinschaft (DFG), among others, under Germany's Excellence Strategy -- EXC 2121 ``Quantum Universe" -- 390833306, and under project number 400140256 - GRK2497; the Hellenic Foundation for Research and Innovation (HFRI), Project Number 2288 (Greece); the Hungarian Academy of Sciences, the New National Excellence Program - \'UNKP, the NKFIH research grants K 131991, K 133046, K 138136, K 143460, K 143477, K 146913, K 146914, K 147048, 2020-2.2.1-ED-2021-00181, and TKP2021-NKTA-64 (Hungary); the Council of Science and Industrial Research, India; ICSC -- National Research Center for High Performance Computing, Big Data and Quantum Computing and FAIR -- Future Artificial Intelligence Research, funded by the NextGenerationEU program (Italy); the Latvian Council of Science; the Ministry of Education and Science, project no. 2022/WK/14, and the National Science Center, contracts Opus 2021/41/B/ST2/01369 and 2021/43/B/ST2/01552 (Poland); the Funda\c{c}\~ao para a Ci\^encia e a Tecnologia, grant CEECIND/01334/2018 (Portugal); the National Priorities Research Program by Qatar National Research Fund; MCIN/AEI/10.13039/501100011033, ERDF ``a way of making Europe", and the Programa Estatal de Fomento de la Investigaci{\'o}n Cient{\'i}fica y T{\'e}cnica de Excelencia Mar\'{\i}a de Maeztu, grant MDM-2017-0765 and Programa Severo Ochoa del Principado de Asturias (Spain); the Chulalongkorn Academic into Its 2nd Century Project Advancement Project, and the National Science, Research and Innovation Fund via the Program Management Unit for Human Resources \& Institutional Development, Research and Innovation, grant B39G670016 (Thailand); the Kavli Foundation; the Nvidia Corporation; the SuperMicro Corporation; the Welch Foundation, contract C-1845; and the Weston Havens Foundation (USA).
\end{acknowledgments}
\bibliography{auto_generated.bib}

\cleardoublepage \appendix\section{The CMS Collaboration \label{app:collab}}\begin{sloppypar}\hyphenpenalty=5000\widowpenalty=500\clubpenalty=5000
\cmsinstitute{Yerevan Physics Institute, Yerevan, Armenia}
{\tolerance=6000
A.~Hayrapetyan, A.~Tumasyan\cmsAuthorMark{1}\cmsorcid{0009-0000-0684-6742}
\par}
\cmsinstitute{Institut f\"{u}r Hochenergiephysik, Vienna, Austria}
{\tolerance=6000
W.~Adam\cmsorcid{0000-0001-9099-4341}, J.W.~Andrejkovic, T.~Bergauer\cmsorcid{0000-0002-5786-0293}, S.~Chatterjee\cmsorcid{0000-0003-2660-0349}, K.~Damanakis\cmsorcid{0000-0001-5389-2872}, M.~Dragicevic\cmsorcid{0000-0003-1967-6783}, P.S.~Hussain\cmsorcid{0000-0002-4825-5278}, M.~Jeitler\cmsAuthorMark{2}\cmsorcid{0000-0002-5141-9560}, N.~Krammer\cmsorcid{0000-0002-0548-0985}, A.~Li\cmsorcid{0000-0002-4547-116X}, D.~Liko\cmsorcid{0000-0002-3380-473X}, I.~Mikulec\cmsorcid{0000-0003-0385-2746}, J.~Schieck\cmsAuthorMark{2}\cmsorcid{0000-0002-1058-8093}, R.~Sch\"{o}fbeck\cmsorcid{0000-0002-2332-8784}, D.~Schwarz\cmsorcid{0000-0002-3821-7331}, M.~Sonawane\cmsorcid{0000-0003-0510-7010}, W.~Waltenberger\cmsorcid{0000-0002-6215-7228}, C.-E.~Wulz\cmsAuthorMark{2}\cmsorcid{0000-0001-9226-5812}
% ...
\par}
\vskip\cmsinstskip
\dag:~Deceased\\
$^{1}$Also at Yerevan State University, Yerevan, Armenia\\
$^{2}$Also at TU Wien, Vienna, Austria\\
% ...
\end{sloppypar}
\end{document}